\documentclass[aip,amsmath,amssymb,reprint]{revtex4-2}

\usepackage{graphicx}
\usepackage{dcolumn}
\usepackage{longtable}
\usepackage{tabularx}
\usepackage{bm}

\usepackage[utf8]{inputenc}
\usepackage[T1]{fontenc}
\usepackage{mathptmx}

\usepackage{amssymb,amsfonts,amsmath}
\usepackage[usenames,dvipsnames]{color}
\usepackage{url}
\usepackage[normalem]{ulem}
\usepackage{cancel}
\usepackage[official]{eurosym}

\usepackage{color}
\usepackage{xcolor}

\begin{document}

\preprint{AIP/123-QED}

\title{Delay-induced homoclinic bifurcations in modified gradient bistable systems and their relevance to optimisation}


\author{Natalia~B.~Janson}
\email[E-mail: ]{N.B.Janson@lboro.ac.uk}
\author{Christopher~J.~Marsden}
\affiliation{Department of Mathematics, Loughborough University, Loughborough LE11 3TU, UK}

\begin{abstract}
Nonlinear dynamical systems with time delay are abundant in applications, but are notoriously difficult to analyse and predict because delay-induced effects strongly depend on the form of the nonlinearities involved, and on the exact way the delay enters the system. 
We consider a special class of nonlinear systems with delay obtained by taking a gradient dynamical system with a two-well ``potential" function and  replacing the argument of the right-hand side function with its delayed version.
This choice of the system is motivated by the relative ease of its graphical interpretation, and by its relevance to a recent approach  to use delay in finding the global minimum of a multi-well function. Here,  the simplest type of such systems is explored, for which 
we hypothesise and verify   the possibility to qualitatively predict the delay-induced effects,  such as a chain of homoclinic bifurcations one by one eliminating local attractors  and enabling the phase trajectory to spontaneously visit vicinities of all local minima. The key phenomenon here is delay-induced reorganisation of manifolds, which cease to serve as barriers between the local minima after homoclinic bifurcations. 
Despite the general scenario being quite universal in two-well potentials, 
the homoclinic bifurcation comes in various versions depending on the fine features of the potential. Our results are a pre-requisite for understanding general highly nonlinear multistable systems with delay. They also reveal the mechanisms behind the possible role of delay in optimisation.

\end{abstract}


\maketitle


\begin{quotation}

Can one predict,  without resorting to numerical analysis, the behaviour of a nonlinear system with time delay described by a delay differential equation (DDE) as the delay is gradually increased? Generally not, as it is 
highly sensitive both to the form of nonlinearities, and to how the delay is introduced. Moreover,  unlike in ordinary differential equations, their phase space is infinite-dimensional, and it is generally impossible to reverse the time, so nonlinear DDEs represent a challenge both for analytical, and for numerical treatment.  However, here we construct a special form of highly nonlinear DDEs, in which one can qualitatively predict a sequence of bifurcations as the delay grows. 
Namely, we slightly modify the most basic setting for optimisation, when the parameter to be optimised decreases as the negative of the gradient of some multi-well ``cost" function  (called ``potential energy landscape function" in physics problems), by delaying  its argument. 
 After overviewing some earlier rigorous results available for simpler DDEs, and interpreting them in terms of the potential function, we use these to predict phenomena in more complex DDEs with two-well potentials, which could be extended to multi-well potentials in the future. 
We hypothesise and verify some universality in the sequence of global homoclinic bifurcations, and also establish how different forms of these bifurcations are realised with different local features of the potential. Delay-induced global bifurcations can be the means to remove the barriers between the local minima of the cost function, and thus to allow the phase trajectory to approach all minima, similarly to  what occurs in a famous optimisation method simulated annealing thanks to random forces. This effect seems promising for optimisation where the delay could replace random forces.  Since the barriers are embodied in the manifolds of saddle points or saddle cycles, their reorganisation via homoclinic bifurcations is key to understanding how the barriers disappear as the delay grows.  We explain rearrangement of manifolds to shed light on this mechanism.

\end{quotation}

\section{Introduction}

Differential equations with time delay represent a special class of dynamical systems, which are routinely used to model the behaviour of  both natural systems and artificial devices alongside with ordinary differential equations (ODEs). Delay equations have been introduced in 1940s as models of population dynamics \cite{Hutchinson_population_dynamis_delay_ANYAS48,Cunningham_delay_population_PNA54} and later of other biological phenomena \cite{Mackey_Glass_oscillation_and_chaos_77,Bocharov_delay_immune_response_JTB94,Cooke_two_delays_SEIRS_epidemic_JMB96,Wolkowicz_delay_chemostat_SJAM97,Culshaw_delay_infection_MB00,Smolen_delay_circadian_oscillators_JN01,Nelson_delay_drug_therapy_HIV_MB02,Ursino_delay_respiratory_AJP03,Villasana_delay_tumor_growth_JMB03,Dhamala_delay_neural_network_PRL04}. In some models, the delay appears in the term(s) \emph{added} to the components(s) of the original ODE \cite{Chow_DDE_added_delay_term_NHMS83,Mallet-Paret_DDE_with_added_delay_term_AMPA86,Chow_DDE_added_delay_term_JDDE89}. Other models contain combinations of delayed and non-delayed terms from the outset \cite{Goodwin_DDE_economics_E51,Cunningham_delay_population_PNA54,Heiden_DDE_chaos_JDE83}.

Delay differential equations (DDEs) present a considerably greater challenge for the analysis than ODEs 
because their state is represented by a (vector-) function on an interval, rather than by a finite-dimensional vector, and hence the dimension of their phase space is infinitely large. Also, time reversal is generally not permitted, which complicates the detection of unstable objects.  Moreover,  the  effects induced by delay greatly depend on a particular form of the DDE under study and on the exact way the delay is introduced. The latter makes it hardly possible to predict, before resorting to numerical analysis and actually observing the behaviour, the dynamics of even a scalar equation with a single delay $\tau$$\ge$$0$, i.e. of $\dot{x}$$=$$f(x,x_{\tau})$ with $x, f$ $\in$ $\mathbb{R}$ and  $x_{\tau}$$=$$x(t-\tau)$ for an arbitrary $f$.  

However, by extending the results from the qualitative theory of ODEs \cite{Arnold_geometrical_methods_book88,Guckenheimer_book97,Kuznetsov_applied_bif_theory_book98,Shilnikov_qual_theory_NLD_book01} it has been possible to qualitatively predict certain phenomena in special cases. Namely,  some predictions were made for DDEs reducible to the form 
$\dot{x}$$=$$f(x_{\tau})$$ -$$ g(x)$, often with $g(x)$$=$$\lambda x$ ($\lambda$$\in$$ \mathbb{R}$) \cite{Chow_DDE_existence_periodic_sols_JDE74,Kaplan_DDE_delayed_non-delayed_JDE77,Heiden_DDE_chaos_JDE83,Chow_DDE_added_delay_term_JDDE89,Wei_DDE_scalar_bifurcation_Nonlin07,Huang_DDE_basins_bistable_JDE14}, or to an even simpler form $\dot{x}$$=$$f(x_{\tau})$, for some special forms of $f$ \cite{Hale_heteroclinic_DDE_JDE86}. 
With this, it is usually impossible to give accurate analytical predictions of the behaviour of general non-linear DDEs, and their studies heavily rely on numerical tools. Even more challenging in DDEs is the detection and interpretation of homoclinic and heteroclinic, i.e. global, bifurcations. 
The reason is that they are based on invariant manifolds, which are usually not confined to a small volume of the phase space (i.e. not localised) and can be one-, two-, many-, or infinite-dimensional.  The tools available to date can handle only \emph{unstable} invariant manifolds in DDEs, and examples considered involve  one-dimensional  
\cite{Green-Krauskopf_unstable_manifolds_DDE_PhD02} and two-dimensional \cite{Sahai_DDE_manifolds_points_cycles_SIAM09,Groothedde_unstable_manifolds_DDE_parametrisation_JCD17} manifolds.

Here, we explore a special class of nonlinear DDEs with bistability, for which we hypothesise and verify the possibility to predict qualitatively how the behaviour changes with the increase of delay. General bistable dynamical systems with delay are popular models in a range of areas, such as optical systems \cite{Gao_optical_bistable_delay_OC86}, laser systems \cite{Green_Krauskopf_laser_bistable_delay_manifolds_PRE02}, neural networks, atmospheric physics \cite{Redmond_delay_bistable_global_bif_PHD02}, and ice dynamics  \cite{Quinn_bistable_delayed_fdbk_DSCS18}, so our work will provide an additional insight into typical bifurcations determining their behaviour. 

The nonlinear DDE to explore is constructed by modifying a nonlinear ODE with arguably the most predictable behaviour, which is  the gradient dynamical system of the form
\begin{equation} 
\label{GDS_nodelay}
 \dot{{x}} =-\nabla V({x}), 
\end{equation} 
where ${x}(t)$$\in$$\mathbb{R}$ is the state, $t$ is the time,
$\dot{{x}}$$=$$\frac{\textrm{d} {x}}{ \textrm{d} t }$, and  $V({x})$$:$$\ \mathbb{R}$$\to$$\mathbb{R}$ is the potential energy function at least twice continuously differentiable  \cite{Hirsch_DS_chaos_book04}.  Also, 
 $\nabla $ is the gradient operator, which for a scalar $x$ is equivalent to $\frac{\partial}{\partial x}$. This system models the behaviour of a particle in a potential energy landscape $V$ immersed in viscous fluid, so that the particle cannot oscillate. Assuming that $V$ is a multi-well function with several maxima separating them, the system converges to one of the local minima in a non-oscillatory manner, and the choice of the minimum depends on the initial conditions.  System (\ref{GDS_nodelay}) represents the most basic setting for optimisation problem, which in practical applications is posed for ${x}(t)$$\in$$\mathbb{R}^N$, $V({x})$$:$$\ \mathbb{R}^N$$\to$$\mathbb{R}$ with  $N$$\ge$$1$, where $V$ is the multi-well ``cost" function, and $x$ is the vector of parameters in need of optimisation \cite{Horst_handbook_global_optimization_95,Pardalos_handbook_global_optimization_v2_02}. Solving this ODE can only deliver a local minimum, so to find the global one this setting is usually  extended  to enable the particle to overcome the barriers between the minima. Most popular extensions rely on incorporating random force that makes the particle visit various regions of the landscape, including the vicinity of the global minimum\cite{Kushner_stoch_optim_book_78,Kirkpatrick_simulated_annealing_Sci83,Aluffi-Pentini_global_opt_stoch_85,Gelfand_stoch_optim_91}.

 We delay the argument of the right-hand side of (\ref{GDS_nodelay}) by some amount $\tau$$\ge$$0$ and thus obtain a DDE to study,
\begin{equation}
\label{dde_whole}
\dot{x}=f(x_{\tau}), \quad f(z)=-\frac{\mathrm{d} V(z)}{\mathrm{d}  z}, 
\end{equation}
where $x$, $f$, $V$, $z$ $\in$ $\mathbb{R}$, $x_{\tau}$$=$$x(t-\tau)$. This form of  (\ref{dde_whole}) allows us to make some rough qualitative predictions about its solutions based on the knowledge of general bifurcation theory and 
of general delay-induced effects in dynamical systems. 
Namely, in 
\cite{Janson_optimization_by_delay_2019}  it was hypothesised  and numerically demonstrated that an increase of $\tau$  could eliminate one by one all local attractors  via global homoclinic   and possibly heteroclinic bifurcations, and give rise to a single large attractor embracing all local minima of $V$. Thus the phase trajectory could be forced to eventually approach all minima of $V$, including the global one, regardless of the initial conditions. 
This means that the delay  could roughly mimic the effect from adding large noise to (\ref{GDS_nodelay}), 
as in global optimisation with simulated annealing \cite{Kirkpatrick_simulated_annealing_Sci83}. However, 
here it would be achieved in a fully deterministic manner  and through an entirely different mechanism.

Specifically, elimination of local attractors is only possible if their individual basins of attraction cease to be separated by boundaries formed by manifolds of saddle fixed points or of saddle cycles. In the context of optimisation, these manifolds serve as the barriers between the minima of $V$. After the homoclinic bifurcations these manifolds do not disappear, but reorganise in such a way, that they cease being the barriers. Thus,  
homoclinic bifurcations could effectively break down the barriers between the local minima. This would provide a mechanism,  alternative to random noise, for overcoming barriers as required in optimisation. 
With this, understanding of the way the manifolds reorganise as the delay grows would provide the key to understanding if and how optimisation by delay could work.  
  
 As a starting point and a pre-requisite for understanding the phenomena in delay systems with general multi-well $V$ depending on one or many variables, here we focus on system (\ref{dde_whole}) with \emph{two-well} smooth landscape functions $V$, such that $V(z)$$\rightarrow$$\infty$ as 
$|z|$$\rightarrow$$\infty$ and $V$, $z$ $\in$ $\mathbb{R}$.

In Section~\ref{DDE_simple} we overview mathematical theorems related to some simpler delay systems, interpret and illustrate them for the general reader, and put into context of a DDE (\ref{dde_whole}) with energy $V$. In Section~\ref{section_eigen} we analyse stability of the fixed points of (\ref{dde_whole}) and its relevance to homoclinic bifurcations.  In Section~\ref{todouble}, 
we combine rigorous theoretical and  quantitative results overviewed in Sections~\ref{DDE_simple}~and~\ref{section_eigen} with the powerful apparatus of the qualitative theory of ODEs \cite{Arnold_geometrical_methods_book88,Guckenheimer_book97,Kuznetsov_applied_bif_theory_book98,Shilnikov_qual_theory_NLD_book01}, which has proved to be successfully applicable to  DDEs  as well, in order to predict, reveal and explain the rather intricate bifurcation phenomena induced by delay in (\ref{dde_whole}) with a double-well $V$. Here, we illustrate these phenomena  with specific examples of  $V$ in (\ref{dde_whole}), demonstrate homoclinic bifurcations of  various types, and reveal distinctions and universality in the delay-induced behaviour  
of such systems. In Section~\ref{sec_opt} we put the delay-induced phenomena in the context of optimisation problem and demonstrate how optimisation could work with two local minima if one uses the delay. 
In Section~\ref{disc} we discuss the results obtained.

\begin{figure}
\includegraphics[width=0.4\textwidth]{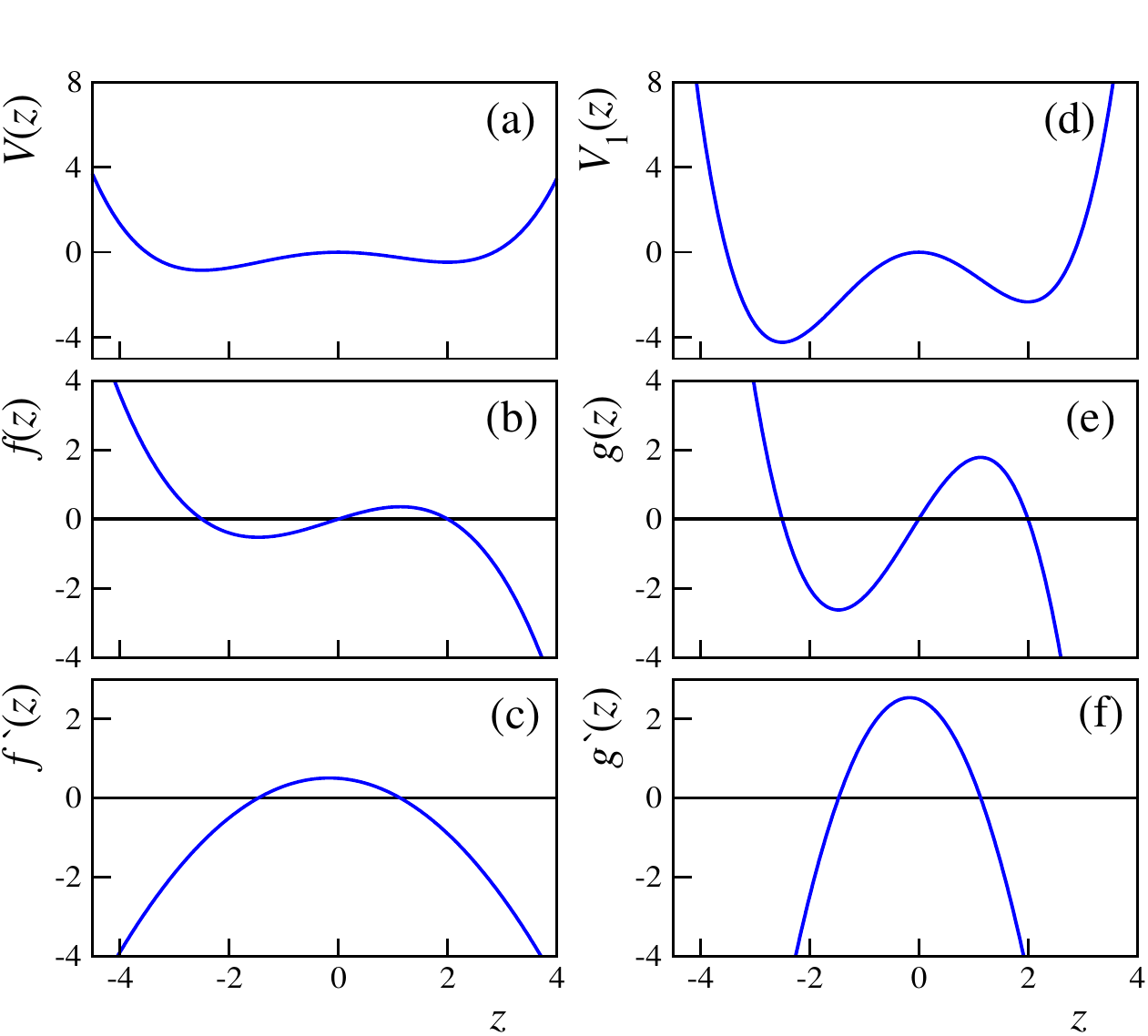}
\caption{Illustration of the relationship between equations (\ref{dde_whole})  with arbitrary $\tau$, and (\ref{dde_tau_1})  with $\tau$$=$$1$ and $g(z)$$=$$\tau f(z)$. For   (\ref{dde_whole}) with $\tau$$=$$5$, left column shows (b) $f(z)$$=$$0.5z-0.05z^2-0.1z^3$, (a) the respective landscape $V(z)$  and (c) $f'(z)$. Right column gives the respective functions for (\ref{dde_tau_1}), namely, (e) $g(z)$$=$$5f(z)$, (d) the landscape $V_1(z)$$=$$5V(z)$  and (f) $g'(z)$$=$$5f'(z)$. }
\label{fig_V_rhs_sample_tau_eq_neq_1}
\end{figure}

\begin{figure}
\begin{center}
\includegraphics[width=0.35\textwidth]{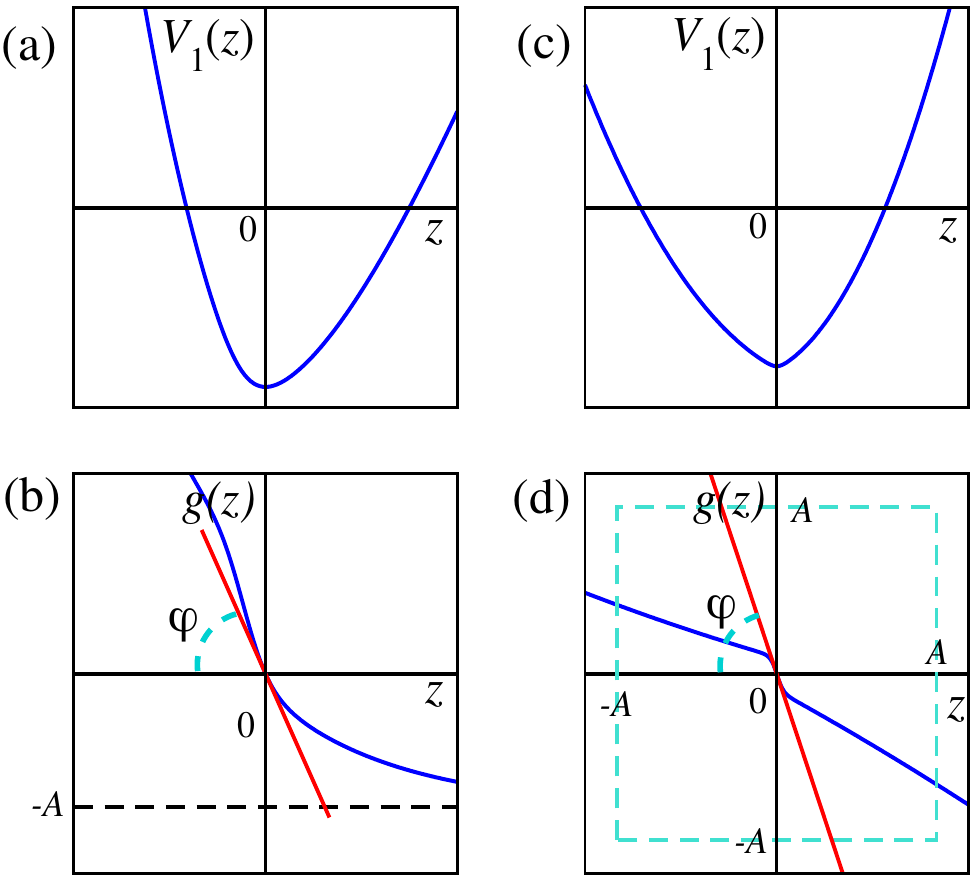}
\caption{Illustration of the theorems for the existence of periodic solutions in (\ref{dde_tau_1}). In (b) and (d) blue lines show two  functions $g(z)$  allowing for periodic solution, and the upper  panels (a) and (c) show the respective landscapes. Periodic solution exists if the slope of the tangent line to $g(z)$ at $z$$=$$0$ (red line)  is $|g'(0)|$$=$$\tan (\varphi)$$>$$\frac{\pi}{2}$. \\ Functions $g(z)$ are: (b) $g(z)$$=$$8\big[ \mathrm{e}^{-(z+z^*)/10}$$-\big(1$$+$$\mathrm{e}^{-(z+z^*)}\big)^{-1}\big]$  with parameter $z^*$$\approx $$1.69132$, such that $g'(0)$$\approx$$-1.72664$, and  \\ (d) 
$g(z)$$=$$\big[$$ -$$\big(1$$+$$\mathrm{e}^{-10z}\big)^{-1}$$+$$0.5\big] \big[ \frac{z}{8}$$+$$1 \big]^2 $$-$$ \frac{2z}{5}$, such that $g'(0)$$=$$-2.9$. }
\label{fig_DDE_tau_1_theorems}
\end{center}
\end{figure}

\section{Delay-induced behaviour in simple systems} 
\label{DDE_simple}

 In order to predict some phenomena that might be caused by delay in systems  (\ref{dde_whole}) with two- and multi-well landscapes $V$, it is important to know about the phenomena occurring in systems with $V$ having only one well. Note that multi-well functions $V$ can be obtained by gluing together the segments of single-well $V$'s and smoothing out all the joints. Thus, bifurcations involving objects localised within a single minimum could be expected in those with more minima, too. This Section overviews the results available for functions $V$ with a single minimum and up to one maximum. 
 
Most theorems fomulated for (\ref{dde_whole}) assume that $\tau$$=$$1$ and $f$ is amendable   \cite{Nussbaum_DDE_periodic_solutions_JDE73,Nussbaum_DDE_periodic_solutions_JDE74,Kaplan_DDE_periodic_JMAA74,Kaplan_DDE_stability_periodic_solution_SIAMJMA75,Mallet-Paret_DDE_JDE76,Nussbaum_DDE_periodic_bif_JMAA76,Chow_DDE_Shilnikov_theorem_homoclinic_TAM89,Walther_DDE_homoclinic_chaos_from_saddle_orbit_NATMA81,Walther_DDE_heteroclinic_to_periodic_TAMS85,Hale_heteroclinic_DDE_JDE86}, i.e. the equation reads
\begin{equation}
\label{dde_tau_1}
\frac{\mathrm{d} x}{\mathrm{d} t}=g(x(t-1)),  \quad g(z)=-\frac{\mathrm{d} V_1(z)}{\mathrm{d}  z}. 
\end{equation}
However, introducing  $t$$=$$s\tau$ and $x(t)$$=$$y(s)$ reduces (\ref{dde_whole}) to 
$\frac{\mathrm{d} y}{\mathrm{d} s}$$=$$\tau f(y(s-1))$, 
 i.e. to (\ref{dde_tau_1}) with $g(z)$$=$$\tau f(z)$ and the appropriate change in notations. Thus, the results valid for (\ref{dde_tau_1}) can be easily adapted to (\ref{dde_whole}) by using the fact that the increase of $\tau$ in (\ref{dde_whole}) is equivalent to sharpening and deepening  the wells of $V_1$ in (\ref{dde_tau_1}), as illustrated in Fig.~\ref{fig_V_rhs_sample_tau_eq_neq_1}.  
 
 Note, that a typical global bifurcation in the DDEs being considered is associated with the formation of a homoclinic orbit, also called a homoclinic loop,  starting and finishing at the same saddle fixed point. For a periodic orbit born from this loop, as the system approaches the bifurcation point, the period tends to become infinitely large. In ODEs there exists another global bifurcation with the same feature of the periodic orbit involved, called saddle-node homoclinic bifurcation and otherwise known as SNIPER, which was first described by Andronov in the 1940s (see \$ 30 of \cite{Andronov_et_al_theory_of_bifurcations_1971}). A SNIPER bifurcation occurs when, as the control parameter changes monotonously, two fixed points (a saddle and a node) move towards each other while staying on the same limit cycle, and disappear in a saddle-node bifurcation. After this bifurcation, the whole of the limit cycle becomes the attractor. In the setting we consider, the only control parameter is the time delay $\tau$, and in DDEs with constant delays the number or locations of the fixed points are not affected by the numerical value of the delay. Thus, the increase of the delay alone cannot lead to the movements or the disappearance of the fixed points, and the SNIPER bifurcation cannot occur in the situations we consider in this paper.

The idea of using delay for optimisation is inspired by the knowledge of typical behaviours of  (\ref{dde_tau_1}) with relatively simple $g$. Although the key relevant theorems are available from highly specialised literature, they have not been previously presented in a form accessible to a more general reader, and  not interpreted in the context of the landscape, which we need before explaining how  we combine them in order to predict the behaviour of (\ref{dde_whole}) with complex-shaped $V$.  In sub-sections below,  three prominent special cases of $g$ in (\ref{dde_tau_1}) are considered, namely, monotonically decreasing with a single zero-crossing,  and making two zero-crossings with a single maximum or a single minimum. We illustrate these cases with specially constructed examples. 

\subsection{Existence of a periodic orbit}
\label{sec_theorem_periodic}
~\vspace{-0.2cm}

In \cite{Nussbaum_DDE_periodic_solutions_JDE73,Nussbaum_DDE_periodic_solutions_JDE74}, Eq.~(\ref{dde_tau_1}) was considered with  $g(z)$ continuously differentiable and monotonically decreasing on some open neighbourhood of $z$$=$$0$, crossing zero at $z$$=$$0$, and in addition satisfying $zg(z)$$<$$0$ for all $z$$\ne$$0$. I.e. $g(z)$ should be strictly positive for negative $z$ and strictly negative for positive $z$. Therefore, the fixed point of (\ref{dde_tau_1}) is at $x$$=$$0$.

In addition, $g(z)$ should either be bounded from below for all $z$ (Fig.~\ref{fig_DDE_tau_1_theorems}(b)), or satisfy $|g(z)|$$\le$$ A$ if $|z|$$\le$$ A$, where $A$ is some positive constant (Fig.~\ref{fig_DDE_tau_1_theorems}(d)). If $|g'(0)|$$>$$\frac{\pi}{2}$, then (\ref{dde_tau_1}) 
has a non-zero periodic solution. In \cite{Kaplan_DDE_stability_periodic_solution_SIAMJMA75} the stability theorem for this periodic solution is proved. From the viewpoint of the landscape, to enable a periodic solution, the single well of the twice differentiable $V_1$ should be sufficiently sharp at the bottom (Fig.~\ref{fig_DDE_tau_1_theorems}(a), (c)). 

Note, that the quoted theorems require that $g(z)$ crosses zero at $z$$=$$0$. 
However, obviously, the same results can be adapted to $g(z)$ crossing zero at any $z$ by making an appropriate change of variable. Thus, while the shape of $g$ contains all the necessary information for the predictions, its localisation on $z$-axis is not essential. 

\subsection{Existence of a homoclinic orbit to a fixed point} 

\label{sec_homo_fixed}

\begin{figure}
\begin{center}
\includegraphics[width=0.35\textwidth]{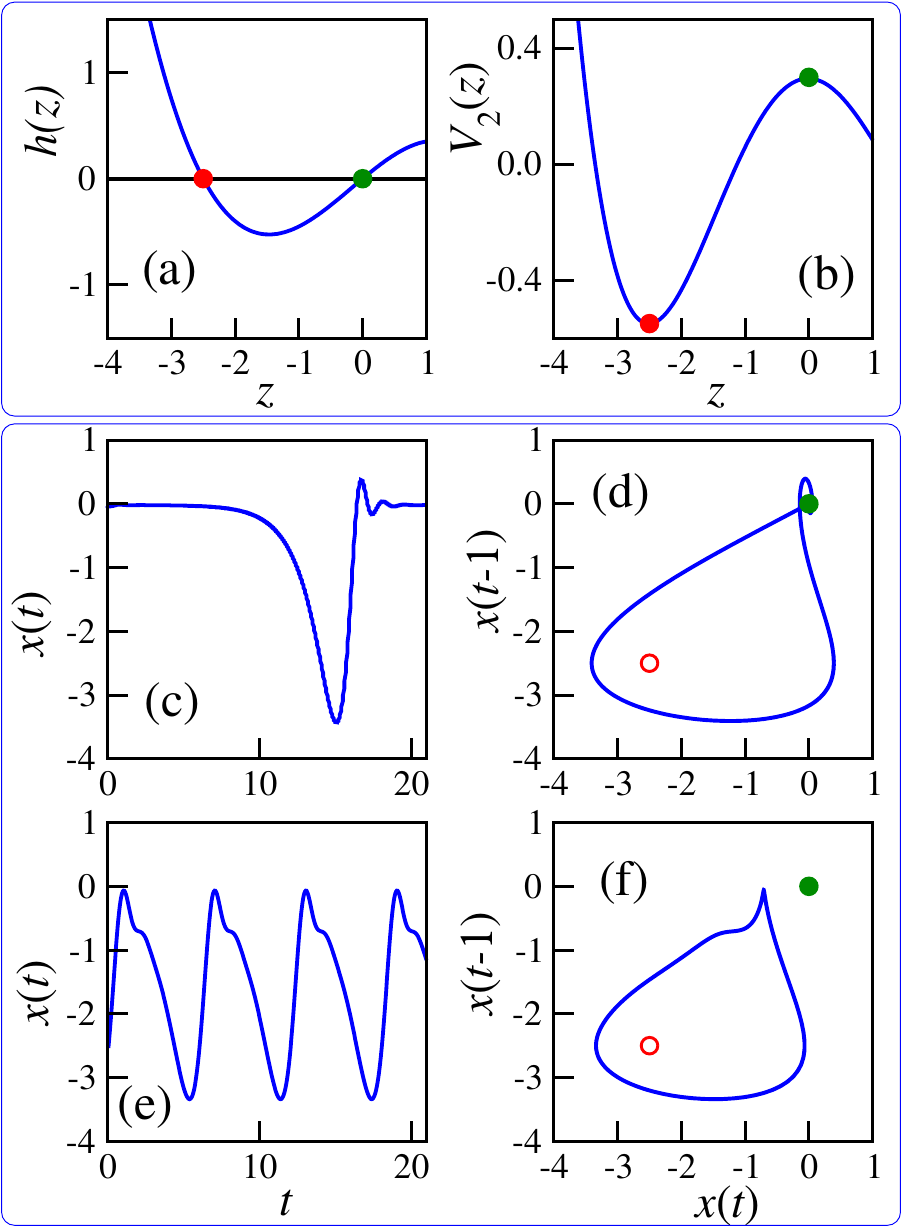}
\caption{Illustration of the theorem for the occurrence of a homoclinic bifurcation in (\ref{DDE_tau_1_homo}), in which a stable periodic orbit collides with the saddle-focus fixed point to form a homoclinic loop. (a) Function $h(z)$$=$$0.5z$$-$$0.05z^2$$-$$0.1z^3$ in a suitable domain $z$$\in$$[-4,1]$, which crosses zero at $z$$=$$0$ and at $z$$=$$-2.5$, (b) landscape $V_2$. Filled circles indicate fixed points: red at the minimum, and green at the maximum of $V_2$. (c) Homoclinic solution existing at $a$$=$$a^*$$\approx$$2.3765$ and (d) the respective phase portrait. (e) Stable periodic solution and (f) the respective orbit at $a$$=$$2.24$.  In (d) and (f) empty red circles indicate the unstable (saddle) fixed point $x=-2.5$ not participating in the homoclinic bifurcation. 
  }
\label{fig_DDE_tau_1_homo}
\end{center}
\end{figure}

Equation  (\ref{dde_tau_1}) with a non-monotonic $g$ was studied in \cite{Walther_DDE_periodic_orbits_from_homoclinic_BCP89,Walther_DDE_Shilnikov_theorem_disser90} but rewritten as

\begin{equation}
\label{DDE_tau_1_homo}
\frac{\mathrm{d} x}{\mathrm{d} t}=ah(x(t-1)), \quad a>0, \quad h(z)=-\frac{\mathrm{d} V_2(z)}{\mathrm{d}  z}. 
\end{equation}
Here, $h(z)$ is required to cross zero at  $z$$=$$0$ from below to above, and at  $z$$=$$-|b|$, $b$$\in$$\mathbb{R}$, from above to below. 
Parameter $a$ controls the sharpness of the two extrema of $V_2$. It has been shown that 
at certain $a$,  in (\ref{DDE_tau_1_homo}) there exists a stable periodic orbit around  $x_2$$=$$-|b|$ (red circle in Fig.~\ref{fig_DDE_tau_1_homo}).   As $a$ increases, this orbit grows in size and, under some additional quantitative conditions on $h(z)$,
at $a$ equal to some critical value $a^*$ can clash with the saddle fixed point at  $x_1$$=$$0$ (green circle in Fig.~\ref{fig_DDE_tau_1_homo}) and form a homoclinic loop. At $a$$>$$a^*$ this orbit no longer exists.

This scenario is illustrated in Fig.~\ref{fig_DDE_tau_1_homo} for a function $h(z)$, which has the required  properties on the domain considered  (see Fig.~\ref{fig_DDE_tau_1_homo}(a) and caption). With this $h$, Eq.~(\ref{DDE_tau_1_homo}) has two relevant fixed points: $x_2$$=$$-2.5$ (red circle) and $x_1$$=$$0$  (green circle)  corresponding to the local minimum and maximum of $V_2$, respectively (Fig.~\ref{fig_DDE_tau_1_homo}(a)--(b)). At small $a$, the point $x_2$ is stable, but it loses stability at $a$$\approx$$1.396$ at which $|ah'(-2.5)|=\frac{\pi}{2}$. The point $x_1$ is unstable for any $a$$>$$0$. At $a$$=$$2.24$ there exists a stable periodic orbit around $x_2$ shown in Fig.~\ref{fig_DDE_tau_1_homo}(f) in projection on the plane $(x(t),x(t-1))$, and the respective solution $x(t)$ is given in (e). At $a$$=$$a^*$$\approx$$2.3765$ a homoclinic orbit is formed as the periodic orbit collides with $x_1$, as shown in (d). In (c) the respective homoclinic solution $x(t)$ is shown, which departs from $x$$=$$0$ starting from some time instant in the past and  approaches $x$$=$$0$ as $t$$\rightarrow$$\infty$ 
\footnote{In ODEs, a homoclinic solution would be called doubly asymptotic to the fixed point (here $x$$=$$0$), which means that it approaches this point when $t$ tends not only to positive, but also to negative, infinity, given that time can be reversed. However, in DDEs  time cannot be generally reversed, so the term \emph{doubly asymptotic} might not be fully appropriate here.}. 

One can interpret these events  in the context of Shilnikov's theorem about the birth of a periodic orbit from the breakdown of the homoclinic loop of a saddle-focus fixed point \cite{Shilnikov_chaos_from_homoclinic_loop_DANSSSR65,Shilnikov_homoclinic_loop_MUSSR68}. 
For $h(z)$ in Fig.~\ref{fig_DDE_tau_1_homo}(b)  for a range of values of $a$$>$$0$, the fixed point $x_1$$=$$0$ of (\ref{DDE_tau_1_homo})  is a saddle focus with a single real positive eigenvalue $\lambda_1$, whereas all of its other eigenvalues are complex-conjugate with negative real parts (see Section~\ref{section_eigen}). 
According to Shilnikov's theorem for ODEs and a relevant study of a special form of   the DDE (\ref{dde_tau_1}) \cite{Walther_DDE_heteroclinic_to_periodic_TAMS85,Walther_DDE_periodic_orbits_from_homoclinic_BCP89},  the breakdown of a homoclinic loop  of a saddle-focus gives birth to a periodic orbit, if at the instant of homoclinic bifurcation the loop is ``safe". The latter means that the saddle quantity  of this saddle fixed point is negative, i.e.  
$\sigma$$=$$\lambda_1+Re(\lambda_{2,3})$$<$$0$, where $\lambda_{2,3}$ are eigenvalues with the negative real parts closest to zero. 

From this viewpoint, in (\ref{DDE_tau_1_homo}) the homoclinic loop shown in Fig.~\ref{fig_DDE_tau_1_homo}(d) is formed as the parameter $a$ 
\emph{decreases} to its critical value $a^*$ from above. As $a$ is further decreased below $a^*$, from this homoclinic orbit a periodic orbit is \emph{born}  (Fig.~\ref{fig_DDE_tau_1_homo}(f)) provided the negativity of $\sigma$ of the fixed point $x_1$$=$$0$. If $\sigma$$>$$0$, then in ODEs, according to Shilnikov, the breakdown of the homoclinic loop should produce chaos rather than periodic behaviour.

\subsection{Existence of chaos from the homoclinic orbit to a saddle cycle} 
\label{sec_homo_cycle}

\begin{figure}
\begin{center}
\includegraphics[width=0.34\textwidth]{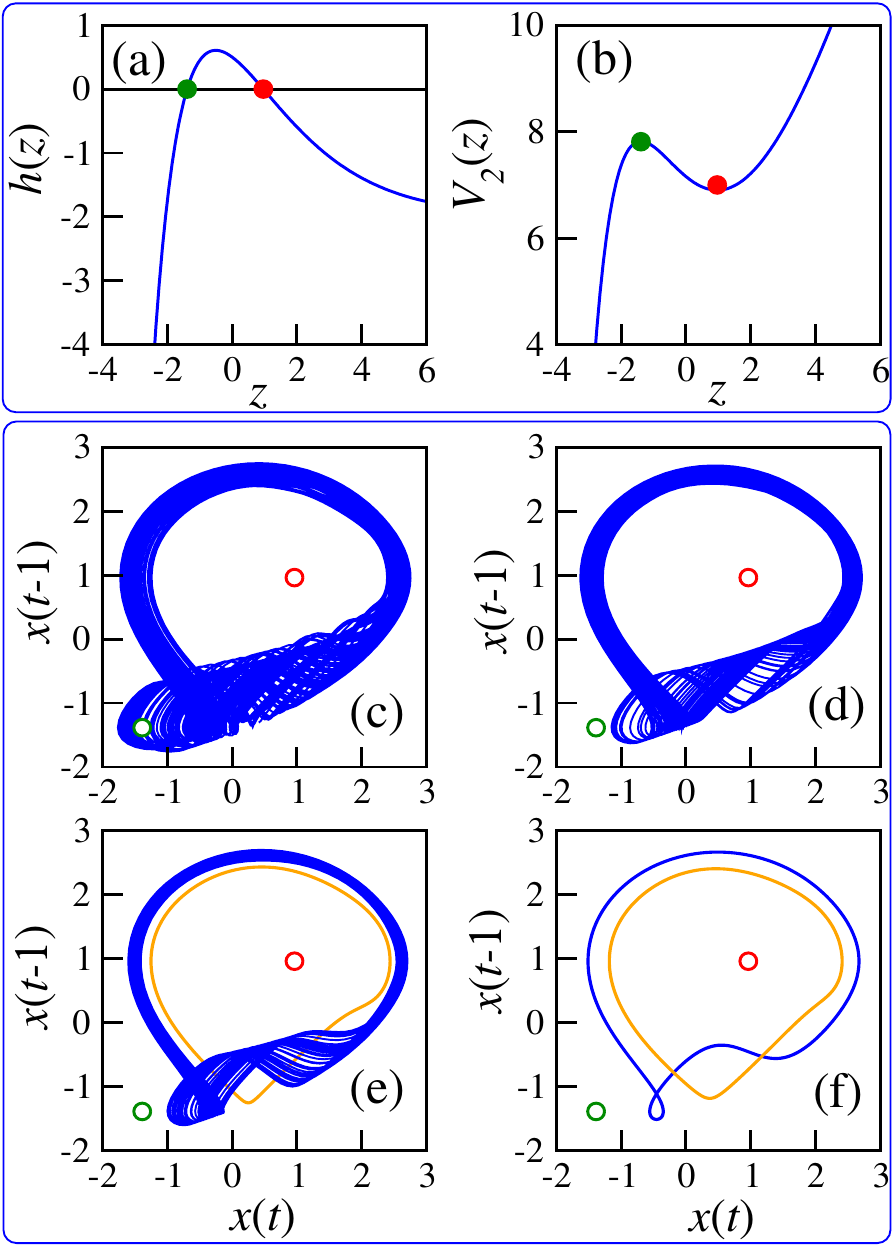}
\caption{Illustration of the theorem for the occurrence of a homoclinic bifurcation  arising from the 
manifolds of a saddle periodic orbit in (\ref{DDE_tau_1_homo}), (\ref{ex_homo_after_AH})  becoming tangent to each other.  Compare  with the bifurcation diagram in Fig.~\ref{fig_DDE_homo_after_AH_BD}.    In (a) $h(z)$ is shown, and in (b) the respective $V_2(z)$. (c)--(f) Phase portraits, with parameter $a$ \emph{decreasing}  from (c) to (f), and   empty circles showing fixed points $x_1$ (green) and $x_2$ (red), which are both saddle at these values of $a$. (c) Chaos from homoclinic bifurcation at the instant of birth at  $a$$=$$3.913$, (d) chaos as the only attractor at $a$$=$$3.85$, (e) chaos  (blue line) coexisting with stable cycle (orange line) at $a$$=$$3.8325$, (f) two stable cycles of different origins coexisting at $a$$=$$3.773$. 
  }
\label{fig_tau_1_homo_after_AH}
\end{center}
\end{figure}

Equation (\ref{dde_tau_1}) with a relatively simple non-monotonic $g$ can demonstrate a different sort of a homoclinic bifurcation arising from 
the  tangency
of a stable and an unstable manifolds of a saddle  (hyperbolic) periodic orbit, which can lead to chaos.  This global bifurcation  was theoretically discovered for ODEs in \cite{Gavrilov_Shilnikov_homoclinic_saddle_cycle_I_Math_USSR_72,Gavrilov_Shilnikov_homoclinic_saddle_cycle_II_Math_USSR_73}, and  described in a more accessible manner in Sec.~7.2.1 of \cite{Kuznetsov_applied_bif_theory_book98}.
In   \cite{Walther_DDE_homoclinic_chaos_from_saddle_orbit_NATMA81,Hale_homoclinic_orbits_DDE_NATMA86}  the existence of such chaos has been proved for very specific DDEs, for which a solution could be constructed analytically. Also, some general properties of function $g$ of (\ref{dde_tau_1}) have been outlined, which should lead to similar behaviour. Namely, $g(z)$ should cross zero twice: from below to above at smaller $z$, and from above to below at larger $z$.  Here we give an example of   a smooth function $g(z)$$=$$ah(z)$ for (\ref{dde_tau_1}), with $h(z)$ expressed as    
\begin{equation}
\label{ex_homo_after_AH}
h(z)=\left( - \frac{1}{1+ \mathrm{e}^{ -\frac{z}{2}}} + \mathrm{e}^{ -\frac{z}{2}} \right) \left(  2-  \mathrm{e}^{ -\frac{z}{2}} \right) 
\end{equation}
and illustrated in Fig.~\ref{fig_tau_1_homo_after_AH}(a), which for $a$$>$$0$ posseses the properties given in \cite{Walther_DDE_homoclinic_chaos_from_saddle_orbit_NATMA81,Hale_homoclinic_orbits_DDE_NATMA86}. The landscape $V_2$, satisfying $h(z)=-\frac{\mathrm{d} V_2(z)}{\mathrm{d}  z}$, is given in Fig.~\ref{fig_tau_1_homo_after_AH}(b).
 In this Section, we study  (\ref{DDE_tau_1_homo}) with $h$ given by (\ref{ex_homo_after_AH}), which is  equivalent to (\ref{dde_tau_1})  with $g(z)$$=$$ah(z)$. We will follow the behaviour of (\ref{DDE_tau_1_homo}),  (\ref{ex_homo_after_AH}) as control parameter $a$ is varied.

\begin{figure}
\begin{center}
\includegraphics[width=0.4\textwidth]{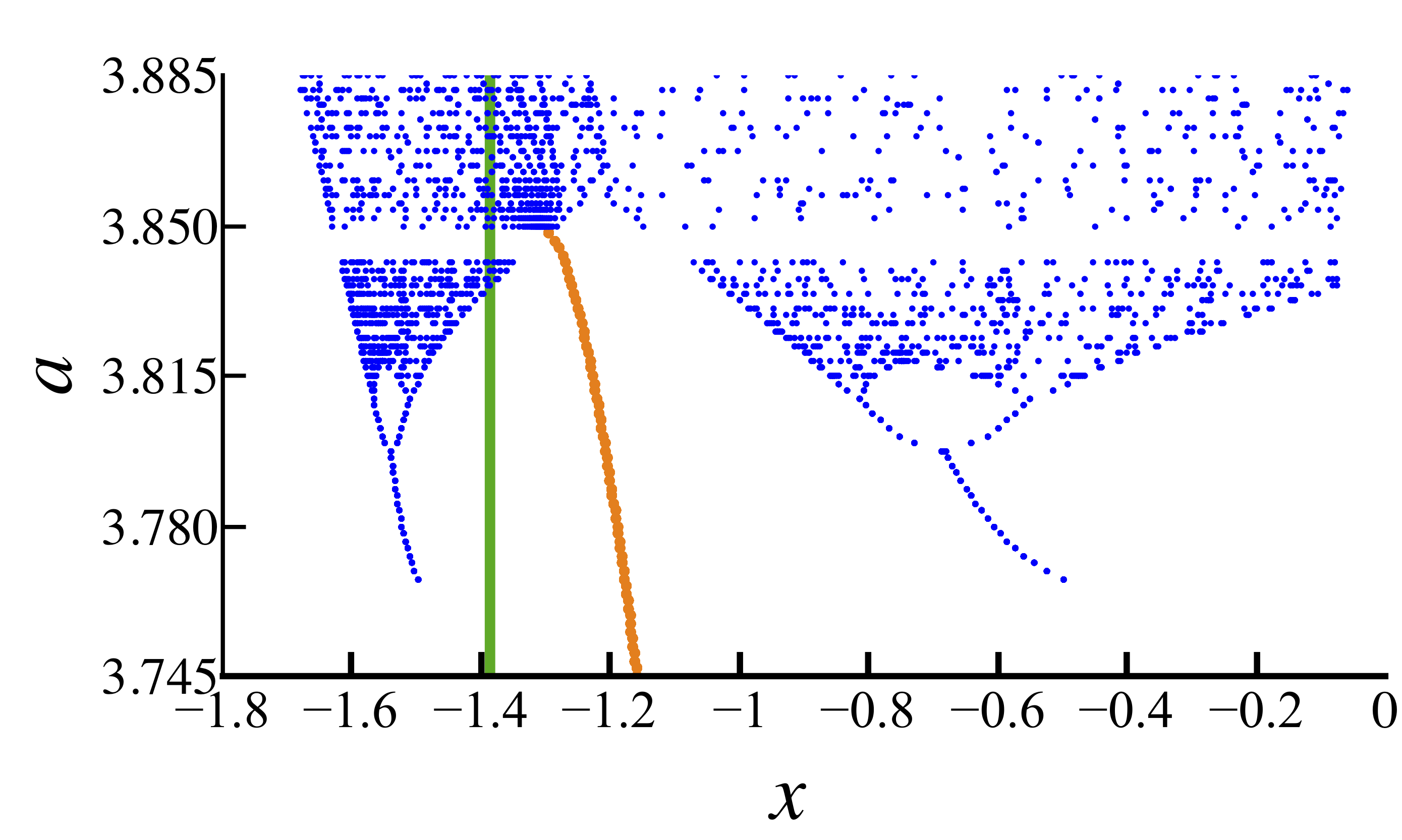}
\caption{A segment of bifurcation diagram of (\ref{DDE_tau_1_homo})  with $h$ given by (\ref{ex_homo_after_AH}),  which for every value of $a$ shows the local minima of the solution $x(t)$   (after transients are removed). Blue dots   correspond to  attractors originating from the homoclinic  orbit to
the saddle cycle located around the saddle fixed point $x_1$ at the maximum of $V_2$ (green line). Orange dots show attractor originating from the fixed point $x_2$ at the minimum of $V_2$, which is outside the range of $x$ of this figure.  One can clearly see bistability for $a$$\in$$[3.77,3.85]$, which means that two attractors of different origins coexist. 
  }
\label{fig_DDE_homo_after_AH_BD}
\end{center}
\end{figure}

 There are two fixed points  here: $x_1$$=$$-1.38629$ (maximum of $V_2$, green circle in Fig.~\ref{fig_tau_1_homo_after_AH}), and $x_2$$=$$0.962424$ (minimum of $V_2$, red circle in Fig.~\ref{fig_tau_1_homo_after_AH}),  which are both saddle for the range of $a$ illustrated in Figs.~\ref{fig_tau_1_homo_after_AH} and \ref{fig_DDE_homo_after_AH_BD}. 
 
 The saddle cycle, whose manifolds eventually  become tangent to each other
 at the homoclinic bifurcation, is born via Andronov-Hopf bifurcation from the saddle point $x_1$ at $a_{AH}$$=$$\frac{1}{h'(z)} \frac{3\pi}{2}$$\approx$$2.8274$ (as explained in Section~\ref{section_eigen}), and exists for $a$$>$$a_{AH}$. This saddle cycle (not shown in Fig.~\ref{fig_tau_1_homo_after_AH}) is located close to its parent $x_1$ (green circle).

  Like in the example of Section~\ref{sec_homo_fixed}, here  the homoclinic  orbit arises from the tangency of
  the manifolds of the saddle cycle  as the parameter $a$ \emph{decreases} to its critical value $a^*  \gtrapprox  3.913 $  from above. The manifolds 
  involved in the formation of this orbit 
are codimension-one (i.e. infinite-dimensional) stable, and two-dimensional unstable, and cannot be visualised with the numerical tools available to date. However, chaos just born from  the manifolds ceasing to be tangent and instead intersecting transversally  
is visualised with blue line in Fig.~\ref{fig_tau_1_homo_after_AH}(c) at $a$$=$$3.913$. 
  
  Generally, in Fig.~\ref{fig_tau_1_homo_after_AH}(c)--(f) phase portraits are shown for (\ref{DDE_tau_1_homo}),  (\ref{ex_homo_after_AH}) for several  values of $a$ as $a$ decreases. Attractors originating from this 
    homoclinic bifurcation 
  are shown by blue line, and attractors originating from the fixed point $x_2$ at the landscape minimum (red cirlce) are shown by orange line. 
  
  The description below can be compared with the bifurcation diagram in Fig.~\ref{fig_DDE_homo_after_AH_BD}. At $a$$=$$3.913$ (Fig.~\ref{fig_tau_1_homo_after_AH}(c))  and $a$$=$$3.85$ (Fig.~\ref{fig_tau_1_homo_after_AH}(d)) chaos born from the homoclinic    bifurcation
  is the only attractor. At $a$$\in$$[3.77,3.84]$ two attractors coexist.  Figure \ref{fig_tau_1_homo_after_AH}(e) shows at $a$$=$$3.8325$  chaos born from the homoclinic  orbit to 
  the saddle cycle  together  with the coexistent stable cycle born through Andronov-Hopf bifurcation from  $x_2$. In 
   Fig.~\ref{fig_tau_1_homo_after_AH}(f) at $a$$=$$3.773$  two coexisting stable cycles are given: the one born from homoclinic chaos in the inverse cascade of period-doubling bifurcations as $a$ decreases    (blue line), and another born from $x_2$ as $a$ increases   (orange line).

\section{Eigenvalues of the fixed points}
\label{section_eigen}

\begin{figure}
\begin{center}
\includegraphics[width=0.45\textwidth]{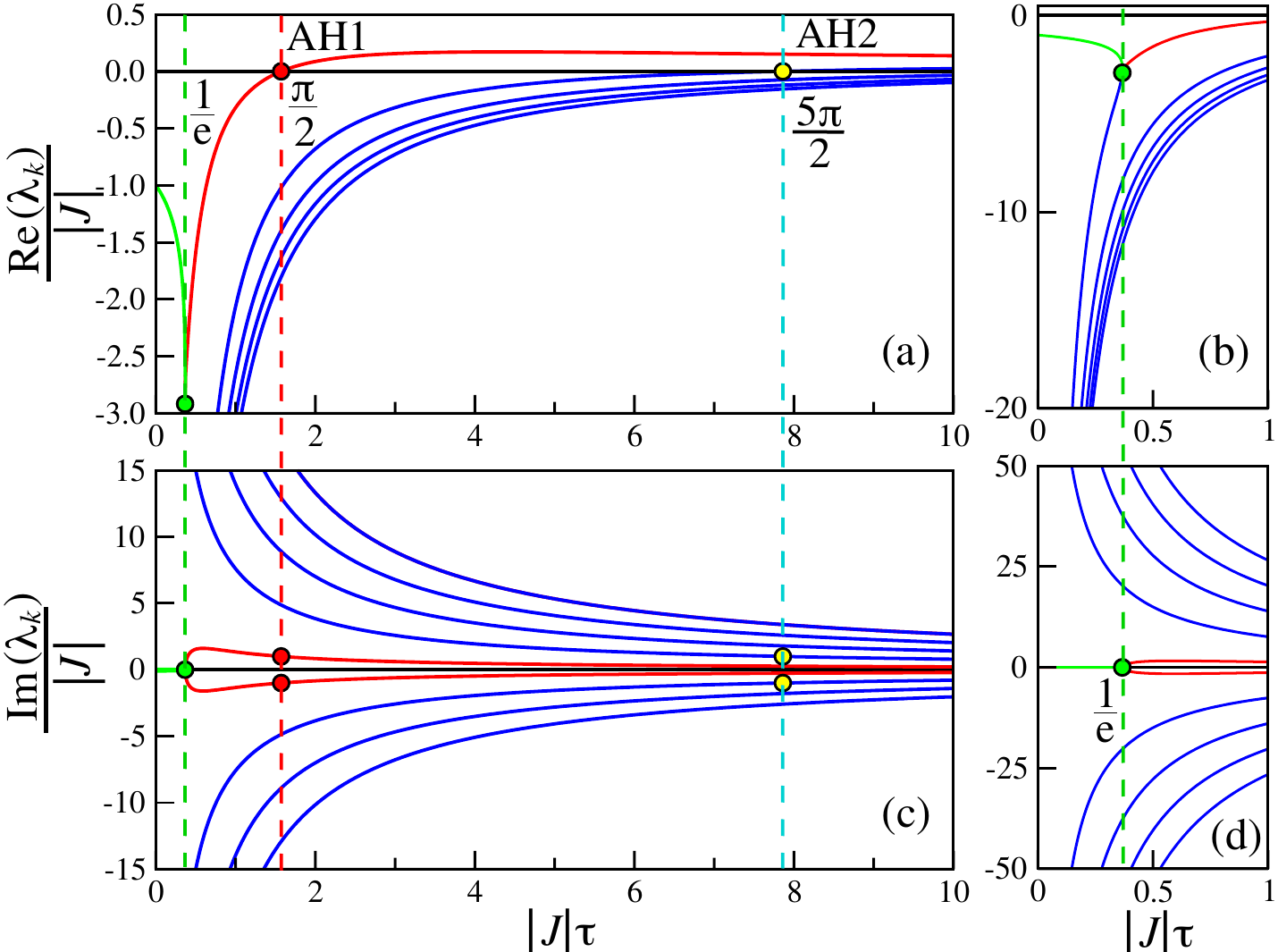}
\caption{(a)--(b) Real  and (c)--(d) imaginary parts of the eigenvalues $\lambda_k$, normalised by $|J|$, of a fixed point corresponding to a minimum of the landscape $V(z)$ in (\ref{dde_whole}) as functions of $|J| \tau$. In (a)--(b) the leading eigenvalue $\lambda_1$ (green line) is real for $|J| \tau$$\in$$ [0,\frac{1}{e})$. 
At $|J| \tau$$>$$\frac{1}{e}$ all eigenvalues are complex. 
At $|J| \tau$$=$$\frac{\pi}{2}$ the real parts of a pair of leading eigenvalues (red line) cross zero signifying the first  Andronov-Hopf (AH) bifurcation marked as AH1 in (a). At $|J| \tau$$=$$\frac{5\pi}{2}$ the second  AH bifurcation takes place, marked as AH2 in (a). Left  and right panels show the same functions in different ranges. }
\label{fig_eigenvalues_min}
\end{center}
\end{figure}

\begin{figure}
\begin{center}
\includegraphics[width=0.35\textwidth]{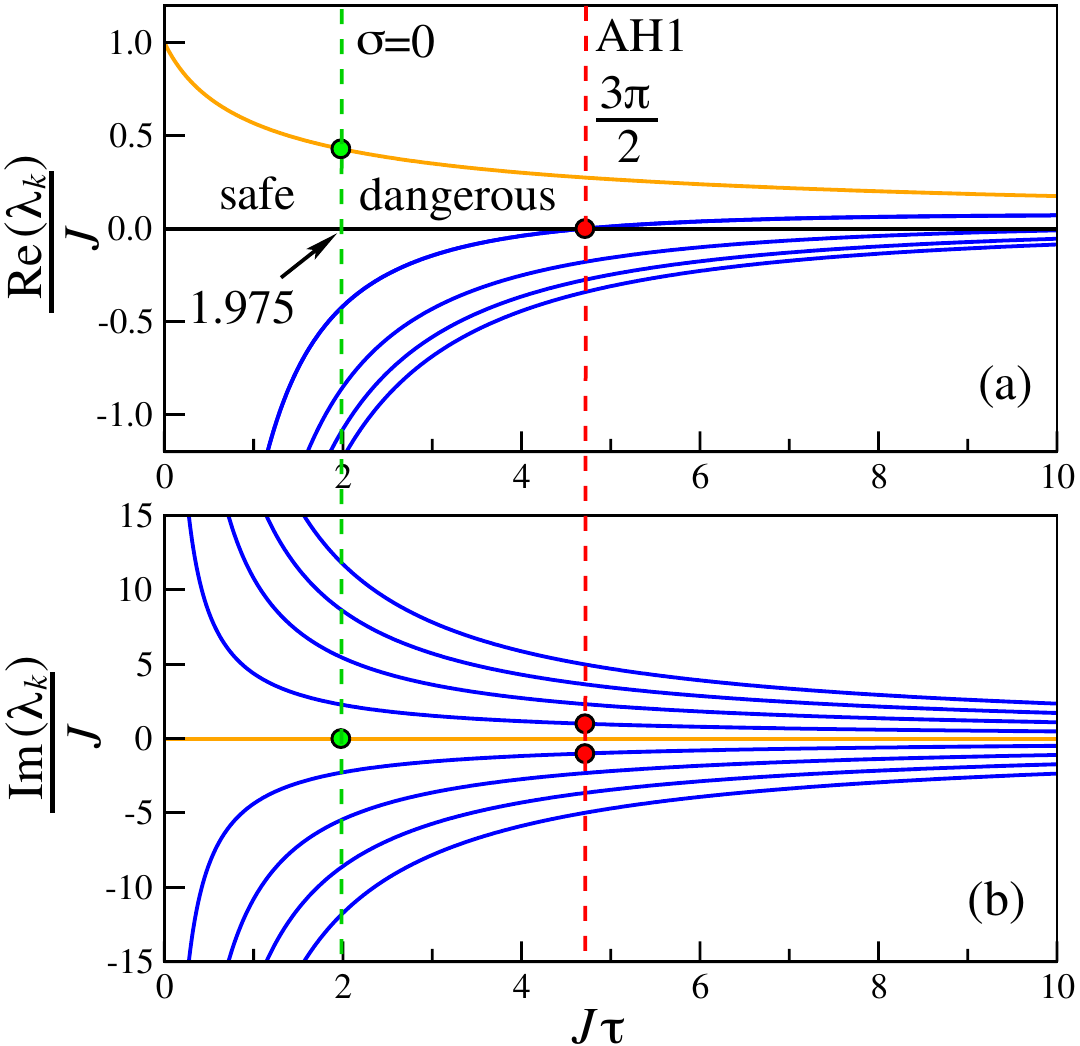}
\caption{(a) Real and (b) imaginary parts of the eigenvalues $\lambda_k$, normalised by $J$,  of a fixed point corresponding to a maximum of $V(z)$ in  (\ref{dde_whole}) as functions of $J \tau$. The leading eigenvalue $\lambda_1$ (orange line) is real for all $\tau$. 
At $J \tau$$\approx$$1.975$ the saddle quantity $\sigma$ changes sign.  At $J \tau$$=$$\frac{3\pi}{2}$ the first  AH bifurcation takes place, marked as AH1.  }
\label{fig_eigenvalues_max}
\end{center}
\end{figure}

From Section~\ref{DDE_simple} it becomes apparent that in (\ref{dde_whole}) with a smooth $V$ having minima and maxima, one should expect an interplay between periodic solutions born from the minima, and attractors born from homoclinic bifurcations associated with the maxima directly or indirectly. To predict the behaviour of the solution near a minimum, or the one resulting from a homoclinic bifurcation related to a maximum, one needs to know the eigenvalues of the relevant fixed points. 
In Section S-I of Supplementary Note we performed the linear stability analysis of the fixed points of (\ref{dde_whole}) and showed that their eigenvalues 
$\lambda_k$ can be expressed in terms of the Lambert function $W(z)$ with $z$$\in$$\mathbb{R}$ and $W$$\in$$\mathbb{C}$ as
\begin{equation}
\label{eig_lambert}
\lambda_k = \frac{W_{k}(J \tau)}{\tau}=\frac{JW_{k}(z)}{z},
\end{equation}
where $J$$=$$f'(x^*)$ and $x^*$ is the fixed point. Function $W(z)$ has countably many branches, as illustrated in Fig.~S2, and  $W_{k}$ is its $k$-th branch. Therefore, $\frac{\lambda_k}{J}$$=$$\frac{W_{k}(z)}{z}$$=$$\frac{W_{k}(J\tau)}{(J\tau)}$. Figures \ref{fig_eigenvalues_min} and \ref{fig_eigenvalues_max} show eigenvalues of the fixed points at the landscape minimum and maximum, respectively (compare with Fig.~S2). 

Note, that at a landscape minimum $J$$<$$0$, so with $\tau$$ \geq$$ 0$, $z$$=$$J \tau$$\le$$0$. However, in Fig.~\ref{fig_eigenvalues_min} for convenience  we show $\frac{\lambda_k}{|J|}$$=$$\frac{W_{k}(-\tilde{z})}{\tilde{z}}$ as a function of  $\tilde{z}$$=$$|J| \tau$$\ge$$0$. 
For $|J| \tau$$\in$$ \big( 0,\frac{1}{e} \big)$ the leading eigenvalue $\lambda_1$ (green line) is real and negative, whereas all other eigenvalues are complex with large negative real parts (this is well visible in Fig.~\ref{fig_eigenvalues_min}(b), (d)). Therefore, the fixed point is effectively a stable node, and the solution converges to it without oscillations. 
At $|J| \tau$$\in$$ \big[ \frac{1}{e},\frac{\pi}{2} \big) $ there is a pair of complex-conjugate leading eigenvalues with negative real parts (red lines in Fig.~\ref{fig_eigenvalues_min}), so in the centre manifold of Andronov-Hopf (AH) bifurcation that occurs at $|J| \tau$$=$$\frac{\pi}{2}$, the fixed point is a stable focus and the solution converges to it in an oscillatory manner. The boundary between these two subtly different types of behaviour is 
$\tau$$=$$ \frac{1}{e}$, which is highlighted by vertical green dashed line in Fig.~\ref{fig_eigenvalues_min}, and the respective values of real and imaginary parts of the eigenvalues are highlighted by green filled circles. 

In Section S-I of Supplementary Note we derive the first condition for the first AH bifurcation  of the fixed point $x^*$ at the minimum  of $V$ (Eq.~(S11)) 
\begin{equation}
\tau_{\textrm{AH1}} = \frac{\pi}{2|J|}, 
\label{AH_min}
\end{equation}
and verify the second condition (Eq.~(S18)). Equation (\ref{AH_min}) is consistent with the predictions for the existence of a periodic solution in (\ref{dde_tau_1}) for a special form of $g(z)$  as discussed in Section~\ref{sec_theorem_periodic}. However, this result is more general  and applies to a smooth $f$ of any shape. Thus,  Eq.~(\ref{AH_min}) allows one to determine the value of $\tau$ at which it is possible for the stable cycle to be born from the fixed point at the landscape minimum. The first AH bifurcation, AH1, is highlighted by the vertical red dashed line in Fig.~\ref{fig_eigenvalues_min}, and the respective values of real and imaginary parts of the eigenvalues are highlighted by red filled circles. At $\tau_{\textrm{AH2}} $$=$$ \frac{5\pi}{2|J|}$ the second AH bifurcation occurs, as a result of which a saddle cycle could be born around the minimum. The second AH bifurcation, AH2, is highlighted with the vertical dashed cyan line in Fig.~\ref{fig_eigenvalues_min}, and the respective values of real and imaginary parts of the eigenvalues are highlighted by yellow filled circles. 

For a fixed point at the maximum of $V$, $J$$>$$0$, and Fig.~\ref{fig_eigenvalues_max} shows $\frac{\lambda_k}{J}$$=$$\frac{W_{k}(J\tau)}{J\tau}$ as a function of $J\tau$$\ge$$0$. One can see that, as also confirmed by the analysis in Section S-I of Supplementary Note, at any $\tau$$\ge$$0$ there is  one real positive eigenvalue $\lambda_1$ (orange line), so this fixed point is always unstable for non-negative $\tau$. However, it is important to appreciate that for 
$J \tau$$\in$$ \big(0,\frac{3\pi}{2} \big)$, this point is a saddle-focus with a one-dimensional unstable manifold and an infinite-dimensional (codimension-one) stable manifold. As shown in (S19) of Supplementary Note, at  $J \tau$$=$$\frac{3\pi}{2}$, or at
\begin{equation}
\tau_{\textrm{AH1}} = \frac{3\pi}{2J},
\end{equation}
the first AH bifurcation occurs, which is highlighted by a vertical red dashed  line in Fig.~\ref{fig_eigenvalues_max}, and the respective values of eigenvalues are marked by red filled circles. For visualisation purposes,  for $J \tau$$\in$$ \big(0,\frac{3\pi}{2} \big)$ we can mentally replace the stable infinite-dimensional manifold of the saddle-focus with a two-dimensional centre manifold of its first AH bifurcation. In this approximation, the given fixed point would represent a saddle-focus with a one-dimensional unstable manifold and a two-dimensional stable manifold,  as explained in more detail in Section~\ref{todouble}.
If there is a well of $V$ nearby, the stable and unstable manifolds can form a homoclinic loop at some $\tau$ from inside $ \big(0,\frac{3\pi}{2J} \big)$. With this, Shilnikov's theorem for ODEs \cite{Shilnikov_chaos_from_homoclinic_loop_DANSSSR65,Shilnikov_homoclinic_loop_MUSSR68} verified for a special form of (\ref{dde_tau_1}) \cite{Walther_DDE_heteroclinic_to_periodic_TAMS85,Walther_DDE_periodic_orbits_from_homoclinic_BCP89}, predicts that  for a safe loop with $\sigma$$<$$0$ (see \ref{sec_homo_fixed} for the definition of $\sigma$ and Shilnikov's theorem), which exists for $J \tau$$\in$$ [0,1.975)$, the homoclinic loop breaks down to form a stable periodic orbit. If the loop is dangerous with $\sigma$$>$$0$, the resultant regime should be chaotic at least in ODEs  \cite{Shilnikov_chaos_from_homoclinic_loop_DANSSSR65,Shilnikov_homoclinic_loop_MUSSR68}, although to the best of our knowledge this was not verified for DDEs. 

Note, that at $ \tau_{\textrm{AH1}}$  a saddle periodic orbit is born from the saddle point at the maximum, which is an intersection of a stable and an unstable manifolds. At larger $\tau$ the manifolds of this saddle orbit can form a loop, whose breakdown can give birth to chaos, as discussed in Section~\ref{sec_homo_cycle}.

\section{Delay-induced behaviour in systems with two-well potentials}
\label{todouble}

A two-well
landscape function $V$ in (\ref{dde_whole}) can be constructed by gluing together segments of single-well landscapes considered in Section~\ref{DDE_simple} and smoothing out the joints. This observation leads us to suggest that the phenomena discussed in Section~\ref{DDE_simple} should also occur in different parts of the phase space of (\ref{dde_whole}) with a  two-well 
$V$. However, a  two-well
function has a different quality as compared to a single-well one, so it is reasonable to also expect new phenomena not covered in Section~\ref{DDE_simple}.

It follows from Section~\ref{DDE_simple} that 
homoclinic bifurcations play a central role in DDEs  (\ref{dde_tau_1})  with non-monotonic $g(z)$ of even  very simple shapes, and result in the disappearance of local attractors when $g$ becomes sufficiently steep. The nature of the homoclinic bifurcations depends on the fine local features of $g$, but their occurrence seems inevitable as the steepness parameter grows. For (\ref{dde_whole}), an equivalent of the steepness parameter is $\tau$, so the homoclinic bifurcations are expected to occur as $\tau$ grows. 

 In this Section we verify our prediction, that homoclinic bifurcations should occur in systems (\ref{dde_whole}) with 
 two-well potentials $V$. We reveal that the occurrence of the particular forms of such bifurcations depends on the individual features of the potential wells and a hump, such as the depth, width and sharpness, as well as on the relationships between them.

We illustrate our findings using two subtly different examples of $V$ which  lead to homoclinic bifurcations of different kinds.  For this purpose,  we construct the functions $V$ in such a way, that by adjusting parameters we could control their local features.   The first example is considered in this Section, and the second one in Supplementary Note, Section S--II. However, our additional studies  with the forms of $V$ specified in  Supplementary Note, Section S--III, which are not reported here due to the lack of space, suggest the generality and reproducibility of the phenomena discovered for the landscapes with the same qualitative features. 

 In relation to optimisation, a particularly important phenomenon at large delays is the disappearance of localised attractors and the existence of the global chaos, which enables the phase trajectory to spontaneously visit the neighbourhoods of all local minima. This effect is similar to the action of random noise on  (\ref{GDS_nodelay}) within simulated annealing \cite{Kirkpatrick_simulated_annealing_Sci83} and superficially might seem quite simple. However, the mechanisms leading to global chaos are quite intricate and involve considerable rearrangements  of manifolds belonging to saddle fixed points or to saddle cycles while the system undergoes a chain of homoclinic bifurcations. It is thanks to these global bifurcations that the barriers separating the vicinities of local minima disappear. Therefore, understanding reorganisation of manifolds is crucial for understanding why and how the delay could help in optimisation. In this Section we explain what happens to manifolds by applying the knowledge about homoclinic bifurcations available from the qualitative theory of ODEs, and verify its predictions with numerical simulations of (\ref{dde_whole}).

\subsection{Model}
\label{sec_model}

\begin{figure}
\begin{center}
\includegraphics[width=0.4\textwidth]{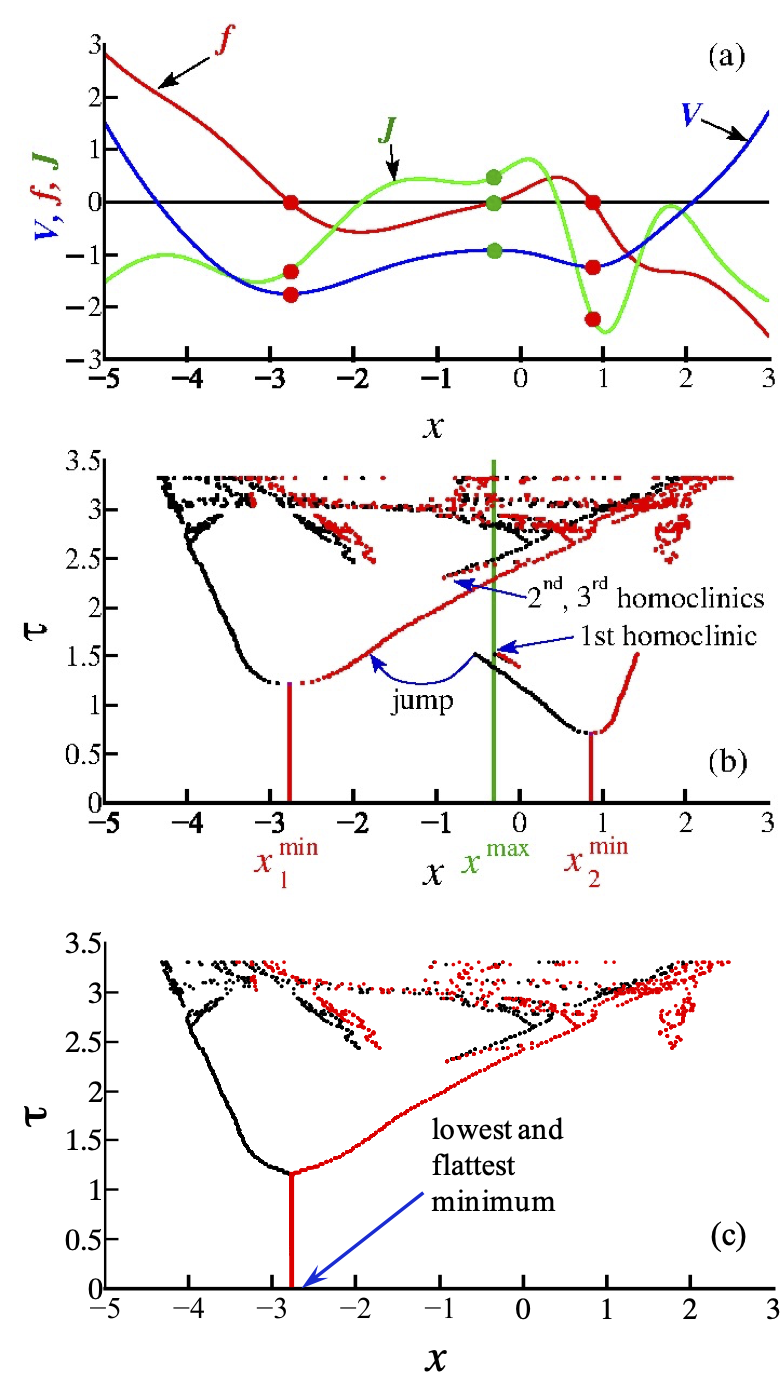}
\caption{(a) Functions $V(x)$ (blue line), $f(x)$ (red line) and $J(x)$$=$$f'(x)$ (green line) specified by (\ref{eq_safe}). Red/green circles show positions of fixed points at the minima/maximum of $V$. \\
(b) Bifurcation diagram of (\ref{dde_whole}), (\ref{eq_safe}). Local minima/maxima of attractors are shown by black/red dots. Fixed points at the minima of $V$ are shown by red vertical lines for $\tau$ at which they are stable, and the saddle-focus at the maximum of $V$ is shown by green vertical line.  As $\tau$ varies, in this system only safe homoclinic loops are formed by the manifolds of the saddle-focus fixed point $x^{\mathrm{max}}$. \\
(c) Demonstration of optimisation. Local maxima (red dots) and minima (black dots) of solutions to (\ref{dde_whole}), (\ref{eq_safe}) are shown, which are obtained as $\tau$ \emph{slowly} decreases from $3.3$ to zero. The solution spontaneously settles down at the lowest, flattest and broadest minimum $x_{1}^{\mathrm{min}}$. Compare with (a) and (b).   }
\label{fig_2well_safe_V_rhs_J_bd}
\end{center}
\end{figure}

\begin{figure}
\begin{center}
\includegraphics[width=0.5\textwidth]{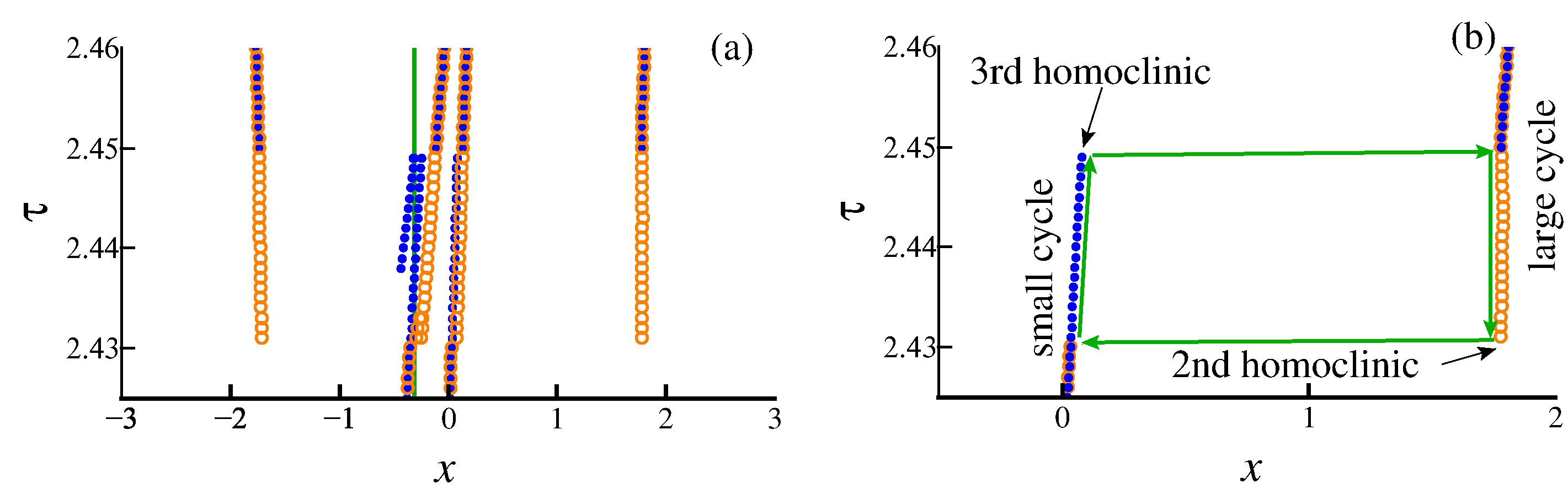}
\caption{(a) Segment of the bifurcation diagram in Fig.~\ref{fig_2well_safe_V_rhs_J_bd}(b) illustrating 2nd and 3rd homoclinic bifurcations, bistability and hysteresis in (\ref{dde_whole}), (\ref{eq_safe}). Circles show all maxima of the attractors (limit cycles): as $\tau$ increases (blue filled), and as $\tau$ decreases (orange empty). Vertical green line shows $x^{max}$. (b) The \emph{largest} maxima of the attractors shown to demonstrate hysteresis more clearly: as $\tau$ increases (blue filled), and as $\tau$ decreases (orange empty). Hysteresis loop is marked by green arrows. The relevant small and large cycles are shown in Fig.~\ref{fig_2well_safe_pp}(h) by red and turquoise lines, respectively. 
}
\label{fig_2well_safe_hom2_left_hyster}
\end{center}
\end{figure}

Here we consider bifurcations in (\ref{dde_whole}) with a double-well $V$ and the respective $f$$=$$-V'$ specified as follows
\begin{eqnarray}
\label{eq_safe}
V(x)&=&-\frac{1}{2} \mathrm{e}^{-2(x-1)^2}- \mathrm{e}^{-0.5(x+3)^2}+0.01(x+1)^4-0.88,  \nonumber \\
 f(x)&=&-2 \mathrm{e}^{-2(x-1)^2} (x-1)-  \mathrm{e}^{-0.5(x+3)^2}(x+3)  \nonumber \\
 &-&0.04(x+1)^3.
\end{eqnarray}
The functions $V(x)$, $f(x)$ and $J(x)$$=$$f'(x)$ are shown in Fig.~\ref{fig_2well_safe_V_rhs_J_bd}(a) by blue, red and green lines, respectively. The constant $(-0.88)$ was added in $V$ in order to make the graphs better distinguishable from each other.  The upper part of Table~I summarises the local features of $V$ in (\ref{eq_safe}), including the positions of two minima $x_{1,2}^{\mathrm{min}}$ and one maximum $x^{\mathrm{max}}$,  their depths $V_{1,2}^{\textrm{min}}$ and $V^{\textrm{max}}$, Jacobians $J_{1,2}^{\textrm{min}}$ and  $J^{\textrm{max}}$, and the points of the first AH bifurcations $\tau_{1,2}^{\textrm{AH1}}$ and $\tau^{\textrm{AH1}}$. In Fig.~\ref{fig_2well_safe_V_rhs_J_bd}(a), red circles indicate the positions of $x_{1,2}^{\mathrm{min}}$,  which are stable fixed points of (\ref{dde_whole}) at $\tau$$=$$0$, and  green circle shows the fixed point at the landscape maximum $x^{\mathrm{max}}$,  which is unstable at $\tau$$=$$0$ and saddle at any  $\tau$$>$$0$.

In the context of optimisation illustrated in Section~\ref{sec_opt}, we remark that in this $V$ the minimum   $x_{1}^{\mathrm{min}}$ is the lowest of the two, and the respective well is the broadest and has the flattest bottom.

\subsection{Overview of bifurcations}

\label{sec_over}

The single control parameter  in (\ref{dde_whole}), (\ref{eq_safe}) is $\tau$, and the bifurcation diagram is shown in Fig.~\ref{fig_2well_safe_V_rhs_J_bd}(b). Namely, vertical red lines show the locations of  fixed points  $x_1^{\mathrm{min}}$ and $x_2^{\mathrm{min}}$ for the range of  $\tau$ inside which they remain stable. At 
$\tau$$=$$\tau_{1,2}^{\textrm{AH1}}$ where the lines stop (values are given in Table~I) AH bifurcations occur in agreement with  (\ref{AH_min}). Green vertical line shows $x^{\mathrm{max}}$ in the whole range of $\tau$ considered. 

For oscillatory attractors, red/black dots indicate local maxima/minima of $x(t)$, which constitute the projections of the Poincar\'e sections defined as $\dot{x}$$=$$0$,  
$\ddot{x}$$<$$0$/$\ddot{x}$$>$$0$, respectively. Note, that this diagram is different from the classical bifurcation diagrams, such as the one in Fig.~\ref{fig_DDE_homo_after_AH_BD}, which usually show only one kind of the attractor extrema. We show both kinds of the extrema because we need to illustrate how the sizes of the limit cycles born from AH bifurcations grow with 
$\tau$, and how the attractors approach the saddle point $x^{\mathrm{max}}$.  This representation allows us to register homoclinic bifurcations from the abrupt changes in the locations and/or amplitudes of the numerically found attractors under very small changes in $\tau$. 

\begin{table}
 \label{tab_1}
\begin{tabular}{| l |}
\hline 
{\bf Features  of the  potential $V$ in (\ref{eq_safe})} \\
\begin{tabularx}{0.45\textwidth}{ X|X|X}
 \hline 
 $x_1^{\textrm{min}}$$=$$-2.77168$ & $x_2^{\textrm{min}}$$=$$0.864388$  & $x^{\textrm{max}}$$=$$-0.308654$  \\  
$V_1^{\textrm{min}}$$=$-1.7557$ $ &  $V_2^{\textrm{min}}$$=$$ -1.2417$ &  $V^{\textrm{max}}$$=$$-0.9207 $ \\
$J_1^{\textrm{min}}$$=$$-1.30015 $  &  $J_2^{\textrm{min}}$$=$-2.19511$ $ &   $J^{\textrm{max}}$$=$$ 0.490368 $  \\
$\tau_1^{\textrm{AH1}}$$=$$1.208 $ & $\tau_2^{\textrm{AH1}}$$=$$0.7155 $ & $\tau^{\textrm{AH1}}$$=$$ 9.61$ \\
flattest, broadest, lowest &  &    $\tau^{\sigma}$$=$4.0276$  $ \\
 \hline \hline
 \end{tabularx} \\
{\bf Homoclinic bifurcation points in (\ref{dde_whole}), (\ref{eq_safe})}  \\
\hline 
 1st homoclinic (safe small loop of  $x^{\textrm{max}}$): $\tau$$\approx$$1.53$, $\tau J^{\mathrm{max}}$$=$$0.75 $\\
 2nd homoclinic (safe large loop of  $x^{\textrm{max}}$): $\tau$$\approx$$2.4307$, $\tau J^{\mathrm{max}}$$=$$1.19$ \\
 3rd homoclinic (safe small loop of  $x^{\textrm{max}}$): $\tau$$\approx$$2.4499$, $\tau J^{\mathrm{max}}$$=$$1.201$ \\
  \hline
\end{tabular}
\caption{ Features of the minima and a maximum of the potential $V$ specified by (\ref{eq_safe}), and the points of homoclinic bifurcations in (\ref{dde_whole}), (\ref{eq_safe}) as $\tau$ grows. 
The Table provides the locations of the minima $x_{1,2}^{\textrm{min}}$ and the maximum $x^{\textrm{max}}$,  their depths $V_{1,2}^{\textrm{min}}$ and $V^{\textrm{max}}$, Jacobians $J_{1,2}^{\textrm{min}}$ and  $J^{\textrm{max}}$, and the points of the first AH bifurcations $\tau_{1,2}^{\textrm{AH1}}$ and $\tau^{\textrm{AH1}}$. For $x^{\textrm{max}}$, $\tau^{\sigma}$ is the value of $\tau$ at which its saddle quantity switches from negative to positive. All  homoclinic bifurcations here are homoclinic loops of the saddle-focus fixed point at $x^{\textrm{max}}$, which are  \emph{safe} since all values of $\tau J^{\mathrm{max}}$ are below $1.975$. 
  }
\end{table}

Altogether there are three homoclinic bifurcations, as indicated in Fig.~\ref{fig_2well_safe_V_rhs_J_bd}(b) and  in Table~I.  Whereas the 1st homoclinic bifurcation can be easily detected from Fig.~\ref{fig_2well_safe_V_rhs_J_bd}(b), the details of the 2nd and the 3rd ones are hard to see in this scale. To demonstrate these convincingly, an enlarged   segment of the bifurcation diagram in the relevant area is provided in Fig.~\ref{fig_2well_safe_hom2_left_hyster}.

\begin{figure}
\begin{center}
\includegraphics[width=0.5\textwidth]{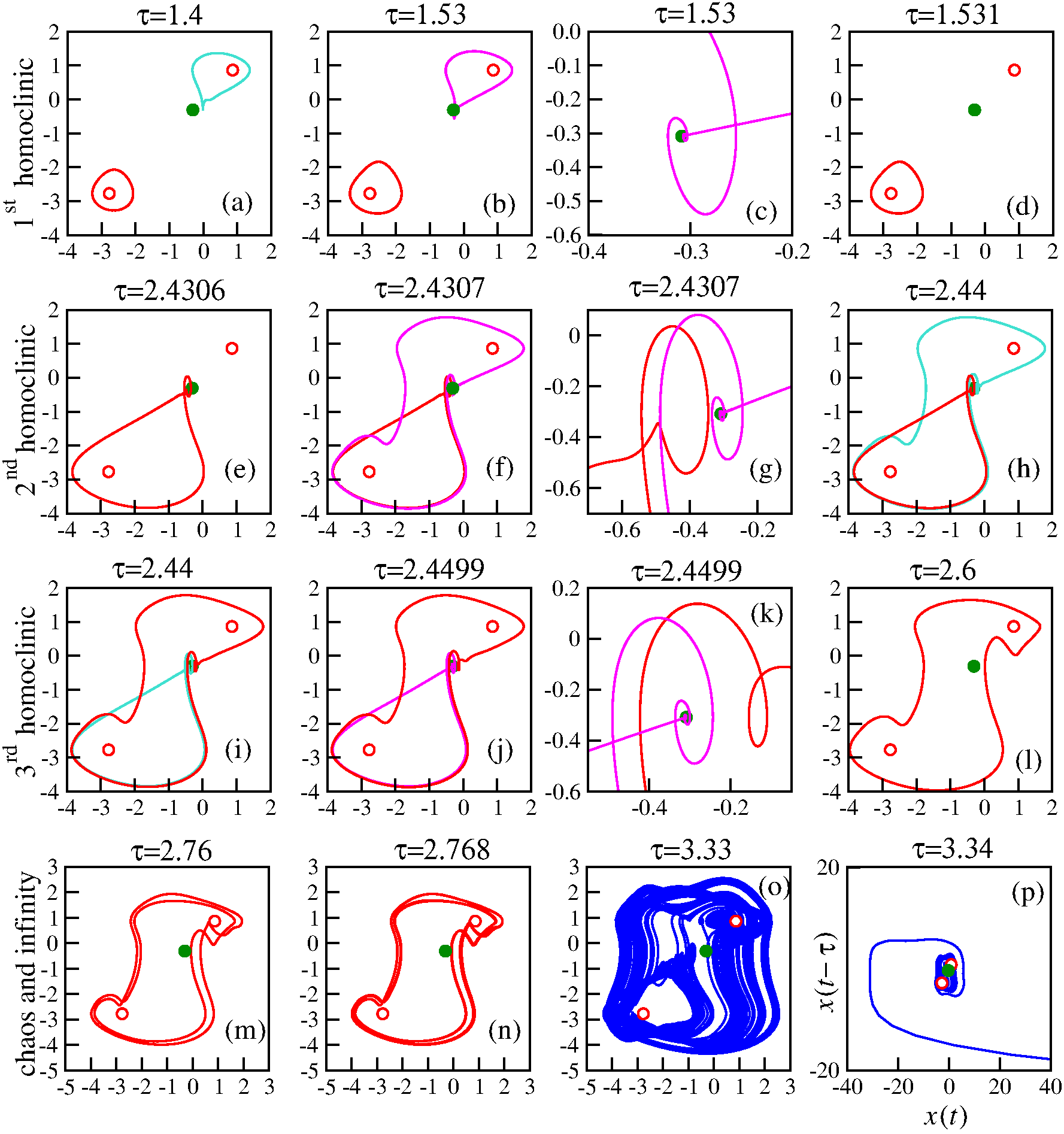}
\caption{Phase portraits of  (\ref{dde_whole}), (\ref{eq_safe}) illustrating bifurcation diagrams in Figs.~\ref{fig_2well_safe_V_rhs_J_bd} and \ref{fig_2well_safe_hom2_left_hyster} with $\tau$ given in panels. Notations are: fixed points $x_{1,2}^{\mathrm{min}}$ (red circles) and $x^{\mathrm{max}}$ (green circle); filled/empty circles indicate points below/above AH bifurcations;  attractors (red and turquoise lines), and homoclinic loops (magenta line). 
{\bf 1$^\textrm{st}$ row:} (a) Before, (b)--(c) at, and (d) after the 1st homoclinic bifurcation. Small cycle around $x_{2}^{\mathrm{min}}$ (turquoise line in (a)) collides with $x^{\mathrm{max}}$  to form a homoclinic loop (magenta line in (b)--(c)) around $x_{2}^{\mathrm{min}}$, and disappears (d).  
{\bf 2$^\textrm{nd}$ row:}
 (e) Before, (f)--(g) at, and (h) after   the 2nd homoclinic bifurcation. (e) Small cycle around $x_{1}^{\mathrm{min}}$ is the only attractor, (f)--(g) homoclinic loop of $x^{\mathrm{max}}$, (h) large cycle born from homoclinic loop (turquoise line) coexisting with small cycle around $x_{1}^{\mathrm{min}}$ (red line).  
 {\bf 3$^\textrm{rd}$ row:}
 (i) Before, (j)--(k) at, and (l) after the 3rd homoclinic bifurcation. (i) Large (red line) and small (turquoise line) cycles coexist, (j)--(k) cycle around $x_{1}^{\mathrm{min}}$ collides with $x^{\mathrm{max}}$ and forms a homoclinic loop, (l) large cycle is the only attractor. 
{\bf 4$^\textrm{th}$ row:}
(m) period-2 cycle, (n) period-4 cycle, (o) Chaos embracing all fixed points at large $\tau$, and (p) trajectory goes to infinity at even larger $\tau$. 
 }
\label{fig_2well_safe_pp}
\end{center}
\end{figure}

Below we will state our predictions about the details of delay-induced homoclinic bifurcations, which can be made on the basis of  rigorous results overviewed in Sections~\ref{sec_homo_cycle} and \ref{section_eigen}. These will be verified and illustrated with phase portraits  in Fig.~\ref{fig_2well_safe_pp}, which complement the bifurcation diagram in Fig.~\ref{fig_2well_safe_V_rhs_J_bd}(b). 

The point $x^{\mathrm{max}}$ is of a saddle-focus type for $\tau $$>$$ 0$. Within the range of $\tau$ shown in Fig.~\ref{fig_2well_safe_V_rhs_J_bd}(b),  
it has a single real positive eigenvalue $\lambda_1$ corresponding to a one-dimensional unstable manifold, and countably many complex-conjugate eigenvalues with negative real parts corresponding to a stable manifold of codimension one.
 In other words, 
$x^{\mathrm{max}}$ does not undergo AH bifurcation and does not give birth to a saddle  cycle. Therefore, we should not be expecting the situation described in Section~\ref{sec_homo_cycle}, and all homoclinic bifurcations here must consist in the closure of the manifolds of  $x^{\mathrm{max}}$ itself to form a one-dimensional loop, following the scenario of Section~\ref{sec_homo_fixed}. 

Based on Section~\ref{sec_homo_fixed}, we can predict that the formation from the loop of a new attractor, which would be localised near $x_{1,2}^{\textrm{min}}$,  would occur as $\tau$ passes the bifurcation point from above to below, i.e. \emph{decreases}. If we consider the sequence of events as $\tau$ is increased, then the localised attractor existing at $\tau$ below the bifurcation point would approach $x^{\mathrm{max}}$, collide with it while the manifolds close at the bifurcation, and cease to exist for $\tau$ above the bifurcation. 

The predicted scenario is confirmed for the 1st and 3rd homoclinic bifurcations illustrated with phase portraits in Fig.~\ref{fig_2well_safe_pp} (a)--(d) and (i)--(l), respectively. As a result of these bifurcations, the localised attractors disappear one after another. However, the 2nd homoclinic bifurcation does not follow from Section~\ref{sec_homo_fixed} and is of a different type. Namely, as $\tau$ grows, this bifurcation produces a large limit cycle embracing all fixed points (see Fig.~\ref{fig_2well_safe_pp}(e)--(h)).

 Figure~\ref{fig_2well_safe_hom2_left_hyster}(a)--(b) shows that in a small range $\tau$$\in$$[2.4307,2.4499]$, i.e. between the 2nd and the 3rd homoclinics, the system demonstrates bistability as the large cycle coexists with the smaller one around  $x_{1}^{\textrm{min}}$, until the latter disappears via the 3rd homoclinic. There is also a hysteresis meaning that  by slowly and continuously increasing and decreasing $\tau$, one observes  different attractors.  Namely,  Fig.~\ref{fig_2well_safe_hom2_left_hyster}(a) shows \emph{all} local maxima of an attractor as $\tau$ increases (blue filled circles) and as $\tau$ decreases (orange empty circles). The lack of the full coincidence between the blue and the  orange circles is an evidence of hysteresis.  Also, a hysteresis loop  is shown by green arrows in (b) where only the largest maxima of the attractors are given for clarity.

 To predict  whether the attractors colliding with $x^{\mathrm{max}}$ at homoclinic bifurcations are limit cycles or chaotic attractors, we need to establish 
whether the respective loops are safe or dangerous, respectively, based on the sign of the saddle quantity $\sigma$, as explained in Sections~\ref{sec_homo_fixed} and \ref{section_eigen}. Namely, as mentioned in Section~\ref{section_eigen}, for DDEs the quantity  $\sigma$ is negative and the loop is safe when $\tau J^{\mathrm{max}}  $$<$$ 1.975$, implying the collision with a stable limit cycle.  For all three homoclinic bifurcations, Table~I gives the values of $\tau$  together with $\tau J^{\mathrm{max}}$, the latter being smaller than $1.975$.  Therefore, we expect safe loops in all three homoclinic bifurcations.  

Phase portraits in Fig.~\ref{fig_2well_safe_pp}(a)--(l) comply with this prediction and show limit cycles being destroyed by ((a)--(d) and (i)--(l)), or born from ((e)--(h)), the homoclinic bifurcations as $\tau$ increases.

\subsection{Role of manifolds in homoclinic bifurcations}

\label{sec_manif}

The key components of homoclinic bifurcations are  invariant manifolds, and it is their reconfiguration which induces most drastic changes in the observed system behaviour associated with the death or the birth of attractors. Here, at $\tau$$>$$0$ different basins of attraction are separated by the {\it stable} manifolds of the saddle-focus $x^{\mathrm{max}}$. Therefore, in the context of optimisation, these stable manifolds form the barriers between the vicinities of the local minima of $V$. 
Therefore, to understand if and how local attractors in (\ref{dde_whole}), (\ref{eq_safe}) can disappear  as $\tau$ grows, one needs to understand how these manifolds are reconfigured.

\begin{figure*}
\begin{center}
\includegraphics[width=\textwidth]{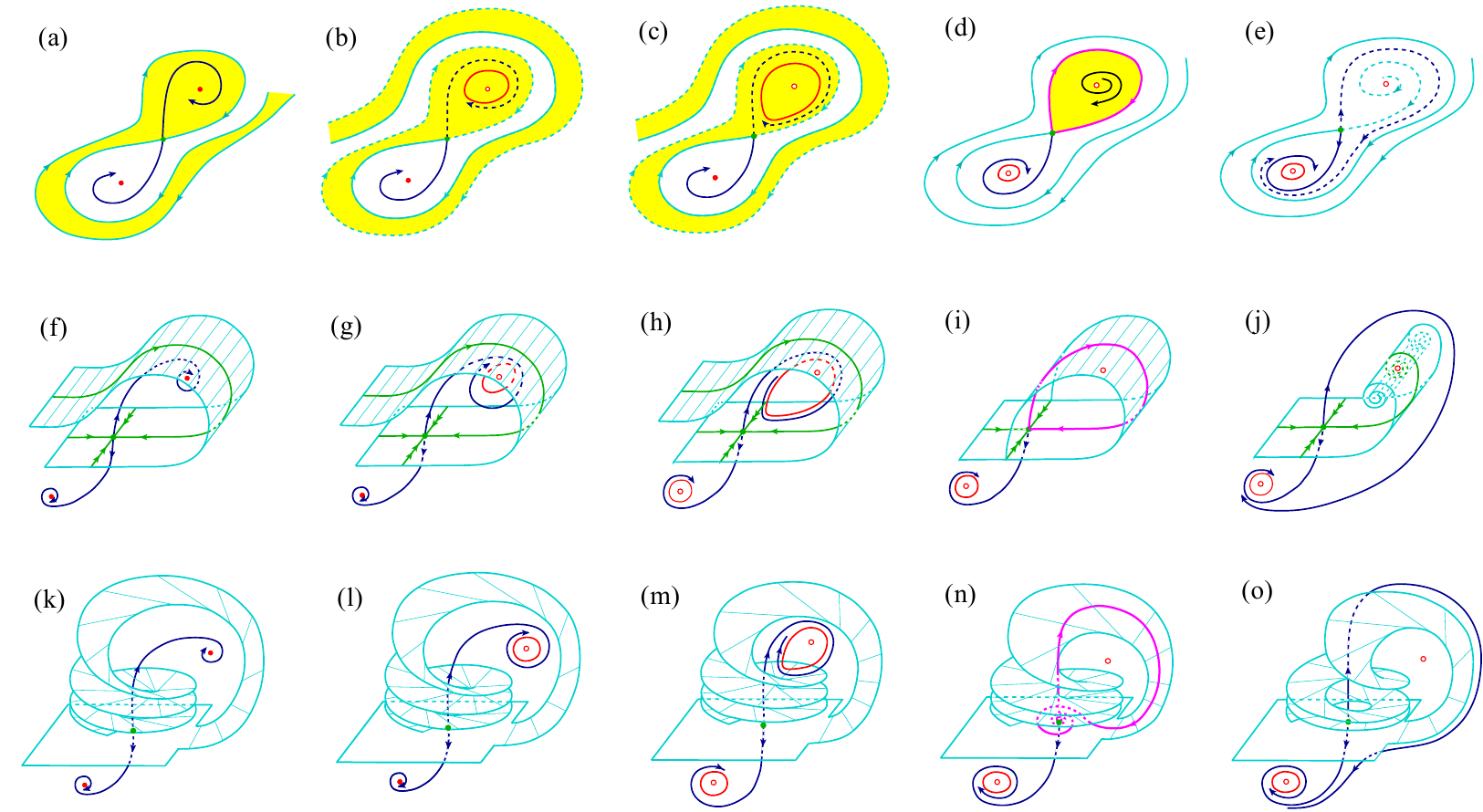}
\caption{Schematic illustration of the 1st homoclinic bifurcation in (\ref{dde_whole}), (\ref{eq_safe})  eliminating the first localised attractor.  \\
 (a)--(j)  Illustrations of a homoclinic bifurcation of a \emph{saddle-node} in some fictional systems of dimensions (a)--(e)  two  and  (f)--(j) three  as some control parameter monotonously changes from (a) to (e), and from (f) to (j), respectively. \\
(k)--(o)  Illustration of a homoclinic bifurcation of a \emph{saddle-focus} $x^{\mathrm{max}}$ in (\ref{dde_whole}), (\ref{eq_safe}), 
 as the value of $\tau$ increases from (k) to (o). \\
Filled red circles show stable fixed points, and empty red circles show fixed points above AH bifurcation, green filled circle shows (a)--(j) a saddle-node and (k)--(o) a saddle-focus, red lines show stable cycles. In (d), (i) and (n) magenta line shows the homoclinic loop. In (a)--(e) yellow and white shades show basins of attraction of two different attractors, and dashed lines in (b), (c), (e) show manifolds involved in homoclinic bifurcation. Cyan/blue lines show stable/unstable manifolds of the saddle-node.  In (f)--(o) cyan surfaces show stable, and blue lines unstable, manifolds of the relevant saddle point. In (f)--(j) green line with double arrows is a {\it strong} stable manifold of the saddle-node, which is contained in its stable 2D manifold, and green lines with single arrows are trajectories on the stable 2D manifold, which approach  the saddle point along its eigenvector  corresponding to the larger of two negative eigenvalues. 
}
\label{fig_homocl_2d-3d_2well_right_cycle}
\end{center}
\end{figure*}

Whereas the attractors of nonlinear DDEs can be detected numerically with relative ease,  revealing and plotting \emph{manifolds} of such systems is a challenge. Some methods to visualise two-dimensional manifolds in nonlinear ODEs have been introduced in \cite{Krauskopf_manifolds_ODEs_CH99,Castelli_ODE_manifolds_2dim_parametrisation_SJAM15}, and  {\it  unstable} manifolds in DDEs  in \cite{Krauskopf_unstable_manifolds_DDE_JCP03,Groothedde_unstable_manifolds_DDE_parametrisation_JCD17}. However,  the phase space of a DDE is infinite-dimensional, implying that the stable manifold of $x^{\mathrm{max}}$ has dimension infinity and is thus highly challenging to detect and to depict. Moreover, revealing a stable manifold in an ODE would require reversing time, which is generally not possible in DDEs \cite{Hale_intro_to_FDE_book93}. With this, to the best of our knowledge, there are no techniques available to date to numerically obtain stable manifolds of DDEs.

\begin{figure*}
\begin{center}
\includegraphics[width=\textwidth]{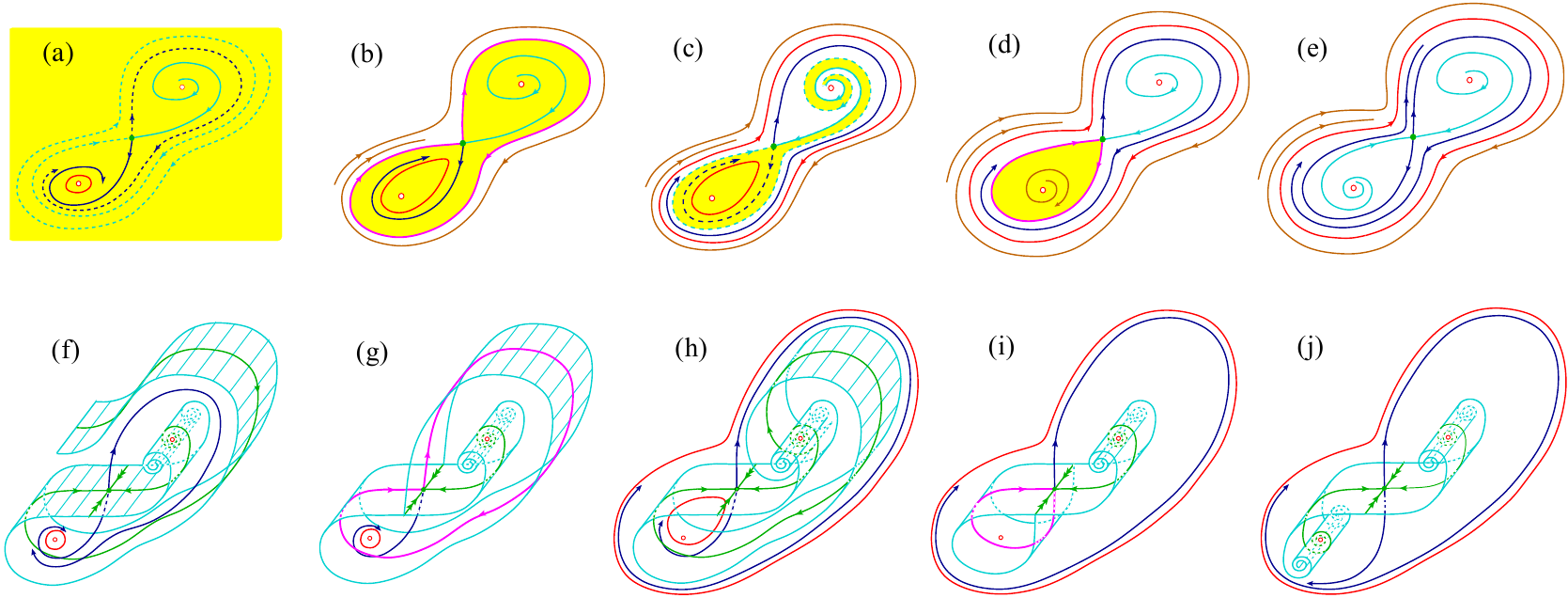}
\caption{Schematic illustration of homoclinic bifurcations in some fictional  (a)--(e) two-dimensional and (f)--(j) three-dimensional  systems,   respectively,  with a saddle-node, which are similar to the 2nd and  3rd homoclinic bifurcations of the saddle-focus in (\ref{dde_whole}), (\ref{eq_safe}) with safe loops forming as $\tau$ is increased from $2.425$ to $2.46$   as shown in Fig.~\ref{fig_2well_safe_hom2_left_hyster}. Notations are as in Fig.~\ref{fig_homocl_2d-3d_2well_right_cycle}, and in addition in (b)--(e) brown line shows a typical phase trajectory. 
 \emph{Away} from the saddle-node, the stable manifold (cyan line or surface) behaves in a qualitatively the same manner as with a saddle-focus (compare (f)--(g) with Fig.~\ref{fig_before+at_large_SF_loop_3D_wide}). \\
 In (b) and  (g) the 2nd homoclinic bifurcation is illustrated, at which a large safe loop (magenta line) is formed by the manifolds of the saddle-node (compare with Fig.~\ref{fig_2well_safe_pp}(f) and Fig.~\ref{fig_before+at_large_SF_loop_3D_wide}(d)). This bifurcation leads to the birth of  a large limit cycle (larger red closed curve in (c) and (h), compare with turquoise curve in Fig.~\ref{fig_2well_safe_pp}(h)). \\
In (d) and (i) the 3rd homoclinic bifurcation is illustrated, at which a small safe loop (magenta line) is formed by the manifolds of the saddle-node (compare with Fig.~\ref{fig_2well_safe_pp}(j)). This bifurcation leads to the disappearance of the small limit cycle and leaves only one attractor being the large limit cycle (red line in (e) and (j), compare with Fig.~\ref{fig_2well_safe_pp}(l)). 
}
\label{fig_homocl_SN_2d-3d_left_cycle_large_cycle}
\end{center}
\end{figure*}

\begin{figure*}
\begin{center}
\includegraphics[width=0.7\textwidth]{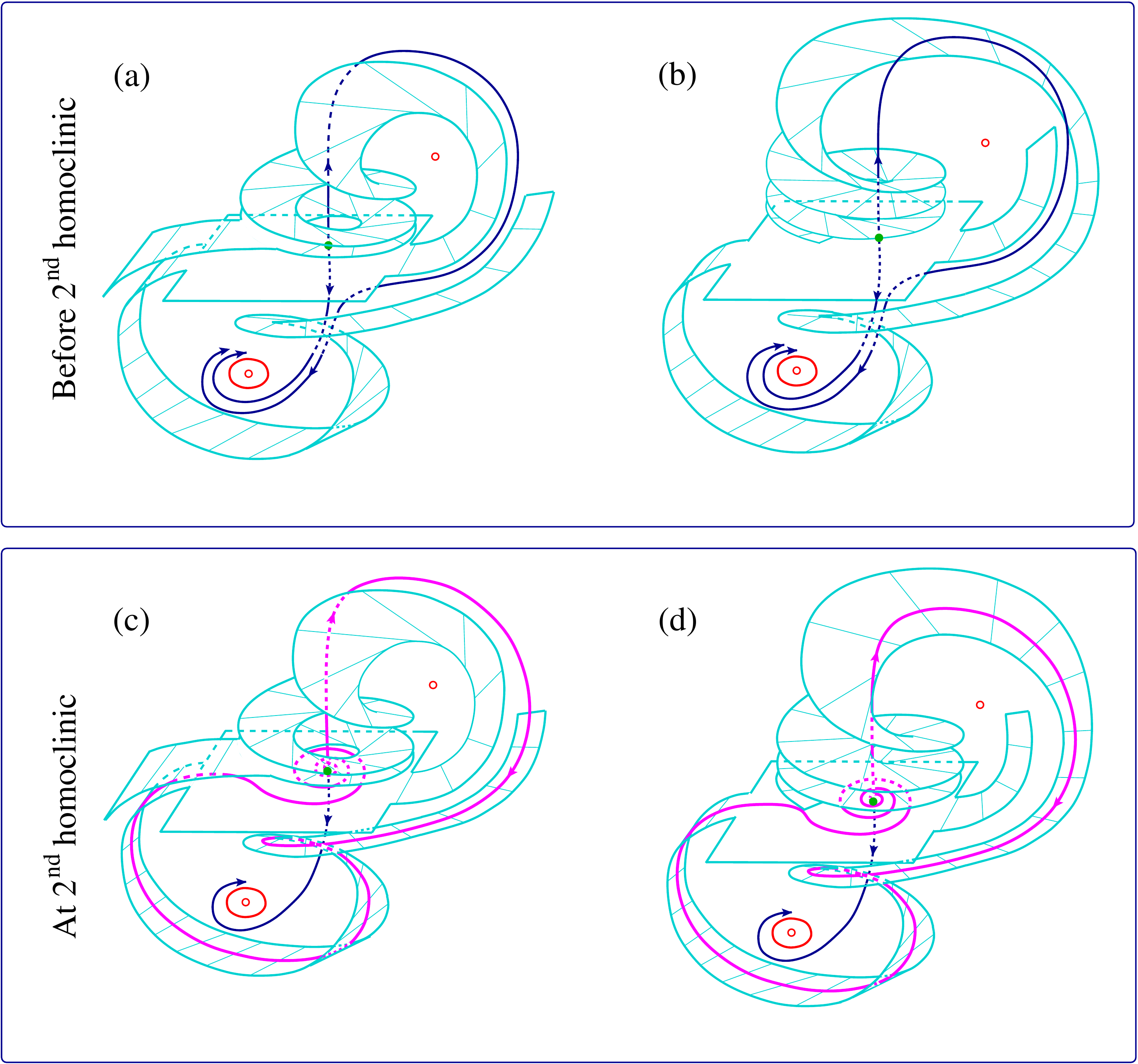}
\caption{Schematic illustration of the 2nd homoclinic bifurcation in (\ref{dde_whole}), (\ref{eq_safe})
 at $\tau$$\approx$$2.4307$, which creates a large safe homoclinic loop of $x^{\mathrm{max}}$ subsequently giving birth to a large  attractor embracing all fixed points. (a)--(b) Before the bifurcation (compare with Figs.~\ref{fig_2well_safe_pp}(e) and  \ref{fig_homocl_SN_2d-3d_left_cycle_large_cycle}(a),(f)), and (c)--(d) at the bifurcation (compare with Figs.~\ref{fig_2well_safe_pp}(f) and  \ref{fig_homocl_SN_2d-3d_left_cycle_large_cycle}(b),(g)).  
Cyan surface shows stable, and blue line shows unstable, manifolds of the saddle-focus at  $x^{\mathrm{max}}$ (green filled circle). 
Unstable fixed points at $x_{1,2}^{\mathrm{min}}$ are shown by empty red circles, and the limit cycle around $x_{1}^{\mathrm{min}}$ is shown by red line in the lower parts of the panels. \\
Spiralling cyan surface in the upper part of each panel shows a portion of the stable manifold (a), (c)  \emph{not} participating in the homoclinic bifurcation, and (b), (d) participating in the bifurcation. The homoclinic loop lies on the portion of the stable manifold shown in (d), which at the bifurcation point and close to $x^{\mathrm{max}}$ has a shape of a closed helicoid with an infinite number of turns. 
}
\label{fig_before+at_large_SF_loop_3D_wide}
\end{center}
\end{figure*}

While being unable to reveal by direct numerical visualisations how the manifolds reorganise  in DDEs, we can still explain this  at a qualitative level by extending the results from the qualitative theory of ODEs \cite{Shilnikov_qual_theory_NLD_book01}, which are supported by numerical visualisations of such reorganisations during homoclinic bifurcations of a saddle-node  \cite{Aguirre_ODE_manifolds_homoclinic_saddle-node_SIAM13} and a saddle-focus \cite{Aguirre_manifolds_Shilnikov_JCD14} in a three-dimensional system of ODEs.  Our explanations and predictions are illustrated with  Figs.~\ref{fig_homocl_2d-3d_2well_right_cycle},  \ref{fig_homocl_SN_2d-3d_left_cycle_large_cycle} and \ref{fig_before+at_large_SF_loop_3D_wide}, which can be compared with the phase portraits in Fig.~\ref{fig_2well_safe_pp}. Although the phase portraits can supply only a highly limited information, their agreement with the predicted effects would provide a reasonably acceptable level of verification for these.

The manifolds of a saddle-focus near a homoclinic loop 
are quite intricate in shape. To make our explanation clearer,  we start from considering some fictional two- and three-dimensional systems, which we assume to demonstrate similar homoclinic bifurcations occurring to a saddle-node, whose manifolds are simpler than those of a saddle-focus (see Figs.~\ref{fig_homocl_2d-3d_2well_right_cycle}(a)--(j) and \ref{fig_homocl_SN_2d-3d_left_cycle_large_cycle}). 

Next, in order to schematically illustrate homoclinic bifurcations in a DDE, we will proceed by analogy with a centre manifold reduction in ODEs \cite{Carr_center_manifold_JDE83}. Namely, we will assume that the dynamics on the infinite-dimensional (codimension-one) stable manifold of the saddle-focus  $x^{\mathrm{max}}$ of (\ref{dde_whole}), (\ref{eq_safe})  can be approximated by the dynamics on the two-dimensional manifold associated with the leading pair of complex-conjugate eigenvalues of this point.  Thus, we will explain reorganisation of manifolds of a saddle-focus in (\ref{dde_whole}), (\ref{eq_safe}) by sketching manifolds, fixed points, limit cycles and  homoclinic loops in a three-dimensional phase space in Fig.~\ref{fig_homocl_2d-3d_2well_right_cycle}(k)--(o) and Fig.~\ref{fig_before+at_large_SF_loop_3D_wide}.

\subsection{First homoclinic bifurcation}

\label{sec_first}

 In Fig.~\ref{fig_homocl_2d-3d_2well_right_cycle}, panels (a)--(e) show snapshots of manifolds,  fixed points and limit cycles at consecutive values of a certain control parameter of a fictional two-dimensional system undergoing the simplest orientable homoclinic bifurcation, which is roughly similar to the 1st homoclinic bifurcation in (\ref{dde_whole}), (\ref{eq_safe}). Here, the saddle-node (green circle) is an equivalent of $x^{\mathrm{max}}$.  Dashed lines show manifolds affected by the homoclinic bifurcation, and solid lines show manifolds unaffected by it. Other notations are given in figure caption.

An equivalent sequence of events in a three-dimensional system is illustrated in Fig.~\ref{fig_homocl_2d-3d_2well_right_cycle}(f)--(j). The only difference from Fig.~\ref{fig_homocl_2d-3d_2well_right_cycle}(a)--(e) is that the stable manifold of the saddle-node is two-dimensional here. Depiction of all manifolds in a 3D space helps to make a transition from a homoclinic loop of a saddle-node to that of a saddle-focus, because the latter can exist only in spaces of dimension three and higher. 

  Next, Fig.~\ref{fig_homocl_2d-3d_2well_right_cycle}(k)--(o) schematically illustrates manifolds of the saddle-focus  and other objects in the phase space of (\ref{dde_whole}), (\ref{eq_safe}):  (k)--(m) before, (n) at, and (o) after the 1st homoclinic bifurcation. The notations are specified in the figure caption.   

The saddle-focus (green filled circle in Fig.~\ref{fig_homocl_2d-3d_2well_right_cycle}(k)--(o)) is an intersection of the two-dimensional stable manifold (cyan surface, a segment is shown) and the one-dimensional unstable manifold (blue line) associated with the only real and positive eigenvalue of this point.  The stable manifold is the boundary between the attractor basins of two attractors (red filled circles or  red lines).

 Below we compare Fig.~\ref{fig_homocl_2d-3d_2well_right_cycle}(k)--(o) with the bifurcation diagram in Fig.~\ref{fig_2well_safe_V_rhs_J_bd}(b) and with the phase portraits in Fig.~\ref{fig_2well_safe_pp}(a)--(d). 
As $\tau$ grows from zero,   the following events precede the 1st homoclinic bifurcation. At $\tau$$\in$$[0,0.7155]$ there are two coexisting stable fixed points (Fig.~\ref{fig_homocl_2d-3d_2well_right_cycle}(k)) at $x_{1}^{\mathrm{min}}$/$x_{2}^{\mathrm{min}}$ below/above the stable manifold of $x^{\mathrm{max}}$  (compare with Fig.~\ref{fig_2well_safe_V_rhs_J_bd}(b)). 

At $\tau$$\approx$$0.7155$, $x_{2}^{\mathrm{min}}$ undergoes AH bifurcation and the stable cycle is born from it. The situation just above this bifurcation is illustrated in Fig.~\ref{fig_homocl_2d-3d_2well_right_cycle}(l), where red line above the stable manifold  shows the newly born cycle. At $\tau$$\approx$$1.208$, $x_{1}^{\mathrm{min}}$ undergoes AH bifurcation and the stable cycle is born from it, which for a slightly larger $\tau$  is given by red line below the cyan surface in Fig.~\ref{fig_homocl_2d-3d_2well_right_cycle}(m) (compare with Fig.~\ref{fig_2well_safe_pp}(a)). With this, the first limit cycle (upper red line in \ref{fig_homocl_2d-3d_2well_right_cycle}(m) and turquoise in \ref{fig_2well_safe_pp}(a)) has grown in size and stretched towards the saddle focus. 

At $\tau$$\approx$$1.53$, the  1st homoclinic bifurcation occurs, namely, the limit cycle around $x_{2}^{\mathrm{min}}$  collides with $x^{\mathrm{max}}$, the unstable manifold of the latter ``sticks" to its own stable manifold and forms a homoclinic loop (magenta lines in Fig.~\ref{fig_homocl_2d-3d_2well_right_cycle}(n) and in Fig.~\ref{fig_2well_safe_pp}(b)--(c)).  This bifurcation is identical to the one illustrated in Fig.~\ref{fig_DDE_tau_1_homo}  as one goes from (f) to (d).  

 As $\tau$ increases slightly beyond the value of the 1st homoclinic $1.53$, the local attractor on the right side of  $x^{\mathrm{max}}$, i.e. near $x_2^{\mathrm{min}}$, ceases to exist. The system has a single attractor left, which is the limit cycle localised near  $x_1^{\mathrm{min}}$ (compare red lines in Fig.~\ref{fig_2well_safe_pp}(d) and Fig.~\ref{fig_homocl_2d-3d_2well_right_cycle}(o)). If at $\tau$ just above $1.53$ the initial conditions are set on or near the just disappeared local attractor around $x_2^{\mathrm{min}}$, the phase trajectory swiftly leaves this region and converges to the limit cycle around $x_1^{\mathrm{min}}$ in what might feel like a ``jump" as indicated in Fig.~\ref{fig_2well_safe_V_rhs_J_bd}(b).

 Unlike in the homoclinic bifurcation of a saddle-node illustrated in Fig.~\ref{fig_homocl_2d-3d_2well_right_cycle}(a)--(j), the shape of the stable manifold of the saddle-focus is quite intricate. Namely,  
close to the homoclinic bifurcation, part of the stable manifold of $x^{\mathrm{max}}$ returns to the vicinity of $x^{\mathrm{max}}$ from above and takes the shape of an open helicoid with a finite number of turns (see upper part of Fig.~\ref{fig_homocl_2d-3d_2well_right_cycle}(k)-(m)), with turns coming closer to each other as they come closer to the saddle-focus, as numerically demonstrated for a 3-dimensional dynamical system in \cite{Aguirre_manifolds_Shilnikov_JCD14}. 

At the instant of homoclinic bifurcation (n), the helicoid becomes closed and develops an {\it infinite} number of turns, such that the distance between the consecutive turns becomes smaller as the turns come closer to $x^{\mathrm{max}}$. At the same instant, the unstable manifold ``sticks" to this closed helicoid and forms a loop (n). After the homoclinic bifurcation, the helicoid opens and possesses a finite number of turns again (o). As a result of homoclinic bifurcation, the one-dimensional unstable manifold (blue line) ``permeates" through the stable one (cyan surface), and the latter no longer  separates basins of attraction.  A more detailed discussion of the attractor basins is given in Section~\ref{sec_basins}.

Figure~\ref{fig_homocl_2d-3d_2well_right_cycle}(k)-(o) schematically shows only a small portion of the stable manifold near the homoclinic loop. However, in reality this manifold has an even more complex shape. E.g. in (o) a part of this manifold should unwind from the vicinity of the unstable fixed point (empty circle in the upper part), but we cannot show this without overloading the figure. Also, this manifold extends well beyond the boundaries shown and has a similarly complex structure in the lower parts of panels in Fig.~\ref{fig_homocl_2d-3d_2well_right_cycle}(k)-(o), which we again do not show for the sake of clarity. Thus, the complexity of the shape of the stable manifold of the saddle-focus prevents us from illustrating this bifurcation in full. 

With this, reorganisation of the manifolds {\it away} from the homoclinic loop during this bifurcation would be qualitatively the same if the saddle-focus is replaced by a saddle node, whose manifolds have a simpler shape and are easier to depict (see Fig.~\ref{fig_homocl_2d-3d_2well_right_cycle}(f)--(j)). 

\subsection{Basins of attraction}

\label{sec_basins}

\begin{figure}
\begin{center}
\includegraphics[width=0.5\textwidth]{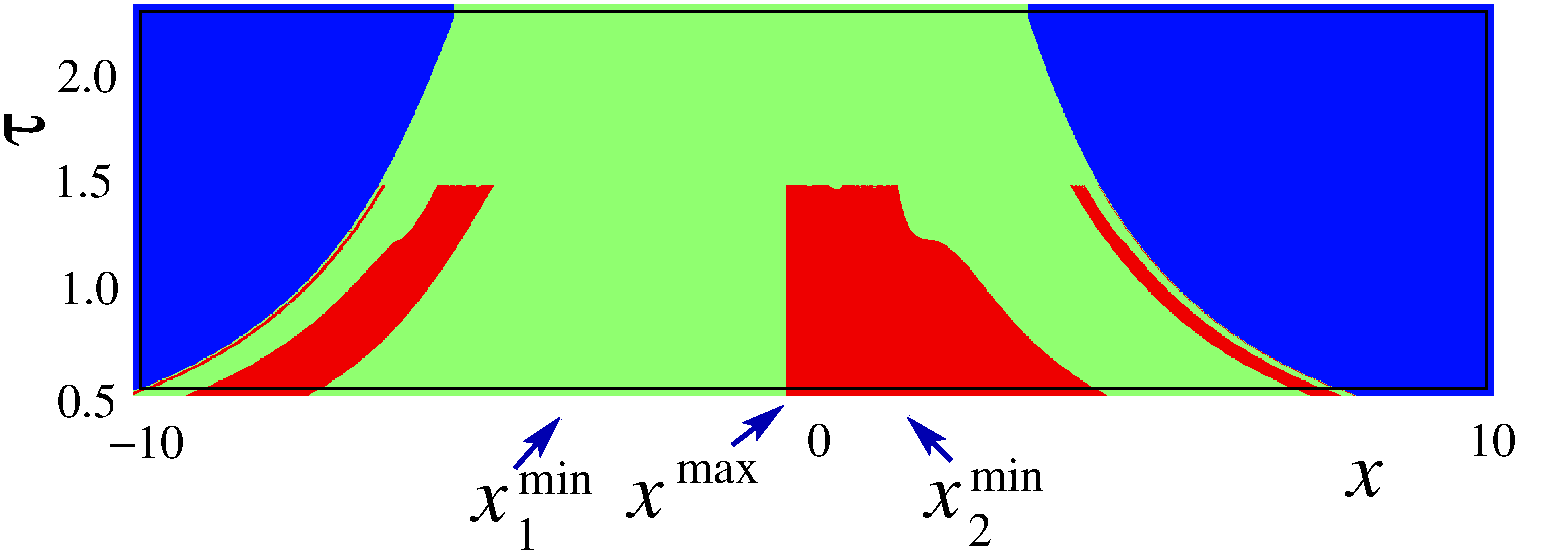}
\caption{Basins of attraction of (\ref{dde_whole}), (\ref{eq_safe}) with safe homoclinic loops, as a function of       $\tau$. Different shades mark  points $x$ such that, from initial conditions  $\varphi(t)$$=$$x$, $t$$\in$$[-\tau,0]$, the phase trajectory converges to 
the following attractors: infinity (blue),  fixed point at, or limit cycle around, $x_{1}^{\mathrm{min}}$ (green), fixed point at, or limit cycle around, $x_{2}^{\mathrm{min}}$ (red). Above  the 1st homoclinic bifurcation point $\tau$$=$$1.53$ there is no local attractor at or near $x_{2}^{\mathrm{min}}$. Compare the stripy structure of the basins below $\tau$$=$$1.53$ here and in  Fig.~\ref{fig_homocl_2d-3d_2well_right_cycle}(a)--(c). 
}
\label{fig_2well_safe_basin}
\end{center}
\end{figure}

In the context of optimisation, our main interest is in how homoclinic bifurcations amend the basins of attraction and lead to their merging.  For a fictional two-dimensional system, Fig.~\ref{fig_homocl_2d-3d_2well_right_cycle}(a)--(e) shows the basins of two coexisting attractors by different shades.  Depiction of the basins is easy on the plane, but difficult in higher-dimensional spaces, so this figure can be used for a visual reference. Here, yellow shade shows the basin of an attractor at or near an equivalent of $x_{2}^{\mathrm{min}}$ (upper right part of the panel), whereas the basin of attractor at or near $x_{1}^{\mathrm{min}}$ (lower left part of the panel) is not shaded. 

 By analogy with similar situations in ODEs, one can hypothesise 
that the stable manifold should make several turns around the three fixed points, thus making both basins of attraction stripy. Although we cannot verify this hypothesis by building  stable manifolds of the DDE, we can numerically find the basins themselves and reveal the stable manifolds as their boundaries,  as was done in \cite{Balanov_saddle_tori_PRE05} for a four-dimensional system.

The basin of attraction in (\ref{dde_whole}) is a set of all initial functions $\varphi(t)$, $t \in [-\tau,0]$, such that all solutions starting from these functions converge to the given attractor. It is a subset  of an infinite-dimensional phase space and difficult to visualise  in full. However, we utilise an approach of \cite{Losson_DDE_basins_Chaos93,Taylor_DDE_chaotic_saddle_PRE07} and consider only a subset of all possible initial conditions, namely, functions of a certain class  parametrised by a finite number of parameters.  In our case we choose the simplest class of constant initial functions, $\varphi(t)$$=$$x$$=$$\mathrm{const}$, $t \in [-\tau,0]$. Such functions can be directly compared with the positions of fixed points   in (\ref{dde_whole}). 

Importantly, numerical simulation of (\ref{dde_whole}), (\ref{eq_safe}), as well as of (\ref{dde_whole}) with several similar potentials,  including those specified in Sections~S--II and S-III of the Supplementary Note, revealed that at positive $\tau$ from some initial conditions such  systems systematically go to infinity. Thus, infinity represents an additional attractor with its own basin and boundaries, which could be revealed numerically.

The basins of attraction  of (\ref{dde_whole}), (\ref{eq_safe}) are given in Fig.~\ref{fig_2well_safe_basin} for a range of $\tau$ in the vicinity of the 1st homoclinic bifurcation. The striking feature of this plot is the large basin of attraction of infinity (blue shade), which expands with growing $\tau$. Below  the bifurcation point $\tau$$=$$1.53$, in addition to infinity, there are two attractors on different sides of $x^{\mathrm{max}}$, and on the plot we do not distinguish  between fixed points and limit cycles.   Due to the 1st homoclinic bifurcation at $\tau$$\approx$$1.53$, the  basin of attractor at or near $x_{2}^{\mathrm{min}}$ (red shade) collapses, and at $1.53$$<$$\tau$$<$$2.43$ besides the infinity there is only one attractor at or near  $x_{1}^{\mathrm{min}}$ (green shade). 

Figure~\ref{fig_2well_safe_basin}  reveals  that the basins of attraction  at least before the 1st homoclinic bifurcation are stripy, and therefore  confirms the hypothesis that the stable manifold of $x^{\mathrm{max}}$ makes several turns around the three fixed points (compare with Fig.~\ref{fig_homocl_2d-3d_2well_right_cycle}(b)--(c)).

\subsection{Second homoclinic bifurcation}

\label{sec_second}

As mentioned in Section~\ref{sec_over}, the 2nd homoclinic bifurcation cannot be predicted based on the theorems overviewed in Section~\ref{sec_homo_fixed}. The numerical simulations suggest that it consists in forming a large homoclinic loop embracing all fixed points (magenta line in Fig.~\ref{fig_2well_safe_pp}(f)), and results in the birth of a large limit cycle (turquoise line in Fig.~\ref{fig_2well_safe_pp}(h)).

The formation of the \emph{large}  homoclinic loop   involves a different pair of manifolds of the saddle-focus $x^{\mathrm{max}}$, as compared to those forming the 1st small loop.
As before, given the difficulty of making clear and easily interpretable sketches of intricately shaped manifolds of a saddle-focus, we initially illustrate the respective homoclinic loop  in a two- (Fig.~\ref{fig_homocl_SN_2d-3d_left_cycle_large_cycle}(a)--(c)) and a three-dimensional (Fig.~\ref{fig_homocl_SN_2d-3d_left_cycle_large_cycle}(f)--(h)) fictional systems with saddle-nodes. Notations are the same as in Fig.~\ref{fig_homocl_2d-3d_2well_right_cycle}, and in addition brown lines in (b)--(e) show typical phase trajectories. 

Figure~\ref{fig_homocl_SN_2d-3d_left_cycle_large_cycle}(a)--(c) explains how the basin of a newly born attractor (large cycle)  is formed thanks to the rearrangement of manifolds. Namely, in (a) yellow shade shows the basin of the only attractor available just before the bifurcation, i.e. of the limit cycle around an equivalent of $x_{1}^{\mathrm{min}}$ (red line, compare with Fig.~\ref{fig_2well_safe_pp}(d)--(e)). At the 2nd homoclinic bifurcation in (b), this basin becomes bounded.  After the bifurcation in (c), the manifolds are rearranged to form the boundary of the new basin (non-shaded) of the large limit cycle born from the homoclinic loop (large red closed curve).  

Figure~\ref{fig_homocl_SN_2d-3d_left_cycle_large_cycle}(f)--(h) demonstrates the same sequence of events, only in a three-dimensional system with a two-dimensional stable manifold of the saddle-node. This illustration is an intermediate stage before depicting manifolds of a saddle-focus in the three-dimensional space while they undergo a similar reorganisation. 

The respective manifolds of a saddle-focus of (\ref{dde_whole}), (\ref{eq_safe}), in the centre manifold reduction and in the vicinity of the  large homoclinic loop, are sketched in Fig.~\ref{fig_before+at_large_SF_loop_3D_wide}. 
Panels (a)--(b) illustrate the situation just before the 2nd (large) homoclinic loop is formed  (compare with Fig.~\ref{fig_homocl_SN_2d-3d_left_cycle_large_cycle}(f)), and (c)--(d) illustrate the instant of this large loop formation (compare with Fig.~\ref{fig_homocl_SN_2d-3d_left_cycle_large_cycle}(g)). 

The picture of the manifolds here is even more complex than near the small homoclinic loop sketched  in Fig.~\ref{fig_homocl_2d-3d_2well_right_cycle}(n) because of the spiralling of the stable manifold as it approaches the saddle focus from \emph{both} sides. However, away from the saddle focus, the stable manifold is qualitatively similar to the one of a saddle-node.

 Before the homoclinic loop, the stable manifold, whose different segments are shown by cyan surface in panels (a) and (b) of Fig.~\ref{fig_before+at_large_SF_loop_3D_wide}, wraps itself around all the three fixed points more than once, thus making more than one layer similarly to the manifold shown in Fig.~\ref{fig_homocl_SN_2d-3d_left_cycle_large_cycle}(a) by dashed cyan line.  

As a result, the one-dimensional unstable manifold of the saddle-focus  (blue line in Fig.~\ref{fig_before+at_large_SF_loop_3D_wide}(a)--(b)) goes \emph{between} the two layers of the stable manifold upwards from $x^{\mathrm{max}}$ and then towards the local attractor near 
 $x_{1}^{\mathrm{min}}$ (red line in the lower left parts of the panels), just like in Fig.~\ref{fig_homocl_SN_2d-3d_left_cycle_large_cycle}(a), (f). At the instant of the homoclinic bifurcation at $\tau$$\approx$$2.4307$, a segment of the stable manifold collides with the saddle-focus and captures the unstable manifold, which now forms a large safe homoclinic loop (magenta line in Fig.~\ref{fig_before+at_large_SF_loop_3D_wide}(d), compare with Figs.~\ref{fig_2well_safe_pp}(f)--(g) and \ref{fig_homocl_SN_2d-3d_left_cycle_large_cycle}(b), (g)). Note, that at the instant of bifurcation, the helicoid-shaped portion of the stable manifold \emph{containing} the homoclinic loop (upper part of Fig.~\ref{fig_before+at_large_SF_loop_3D_wide}(d)) makes an infinite number of turns as it approaches the saddle focus. 
 
 As $\tau$ grows, the loop disappears and gives birth to a large limit cycle embracing all three fixed points,  which are similar to large red closed curves  in Fig.~\ref{fig_homocl_SN_2d-3d_left_cycle_large_cycle}(c), (h). The respective large cycle of (\ref{dde_whole}), (\ref{eq_safe}) is given by turquoise line in Fig.~\ref{fig_2well_safe_pp}(h). 

\subsection{Third homoclinic bifurcation, chaos and infinity}

\label{sec_third}

The 3rd homoclinic bufurcation takes place at  $\tau$$=$$2.4499$ (see Fig.~\ref{fig_2well_safe_V_rhs_J_bd}(b)) and  eliminates the small limit cycle to the left of $x^{\mathrm{max}}$, i.e. around $x_{1}^{\mathrm{min}}$, as illustrated with phase portraits in Fig.~\ref{fig_2well_safe_pp}(i)--(l) and with sketches of a similar bifurcation in systems with a saddle-node in Figs.~\ref{fig_homocl_SN_2d-3d_left_cycle_large_cycle}(c)--(e) and (h)--(j). This  bifurcation is qualitatively the same as the 1st homoclinic. At $\tau$$>$$2.4499$ there are no more local attractors in the system. 

At $\tau$$\in$$(2.4499, 3,33]$, the system has a single large attractor enclosing all fixed points. This attractor is initially a limit cycle of period one (Fig.~\ref{fig_2well_safe_pp}(l)), but with increasing $\tau$ it undergoes a cascade of period-doubling bifurcations (Fig.~\ref{fig_2well_safe_pp}(m)--(n)) and becomes chaotic ((Fig.~\ref{fig_2well_safe_pp}(o)). Note, that the phase trajectory on the  chaotic attractor visits the close vicinities of all three fixed points,  which is roughly similar to the effect caused by adding large random noise to (\ref{GDS_nodelay}) in simulated annealing. 

As $\tau$ exceeds $3.33$, the chaotic attractor disappears and the trajectory goes to infinity (Fig.~\ref{fig_2well_safe_pp}(p)).  We cannot specify the exact reason for this, and can only hypothesise that the manifold bounding the basin of attraction of chaos appears involved in some global bifurcation. Since we cannot visualise this manifold, we cannot verify our hypothesis. However, our studies of a considerable number of systems of the form (\ref{dde_whole}) with various multi-well landscapes $V$ suggest the universality of this phenomenon. 

Namely, it seems that in such systems, as $\tau$ becomes sufficiently large, a chaotic attractor enclosing all extrema of $V$ is  initially born (although it might turn into a periodic one at even larger $\tau$  e.g. when periodic windows in chaos appear \cite{Strogatz_chaos_book94}). However,  when $\tau$ is increased further, the large attractor is inevitably destroyed and the trajectory escapes to infinity from all initial conditions. 

It is important to appreciate that the infinity is present as some kind of an attractor at all values of $\tau$$>$$0$.  To make more specific predictions about encountering attractors at infinity,  a rigorous investigation of the necessary and sufficient conditions for the existence of unbounded solutions in nonlinear systems of type (\ref{dde_whole}) would be required.

\subsection{Different forms of homoclinics}

We also studied the phenomena induced by the increase of $\tau$ in a slightly different system of the form (\ref{dde_whole}) with a double-well landscape $V(x)$  specified by Eq.~(S21) in Supplementary Note, in which at the instants of homoclinic bifurcations the saddle focus was below AH bifurcation, but had a positive saddle quantity. According to Shilnikov's theorem for ODEs \cite{Shilnikov_chaos_from_homoclinic_loop_DANSSSR65,Shilnikov_homoclinic_loop_MUSSR68}, in such cases the homoclinic loops are expected to be dangerous, and their breakdown should lead to the formation of chaotic attractors. However, to the best of our knowledge, this result has not been verified for DDEs. Our numerical studies did not reveal any obvious differences between the sequence of bifurcations in the systems of the form (\ref{dde_whole}) with double-well landscapes $V$ with dangerous or safe homoclinic loops, and the observed phenomena looked very similar when studied with the same numerical accuracy. Thus, whether the dangerous homoclinic loop gave birth to chaos or not, it did not affect the order of bifurcations, and the same key phenomena were observed.

In addition, Section~S--II of the Supplementary Note presents a similar, albeit briefer, analysis of bifurcations in (\ref{dde_whole}) with a two-well potential $V$, for which at the instant of the 1st homoclinic bifurcation the fixed point $x^{\mathrm{max}}$ is past AH bifurcation.  There, the homoclinic   orbit 
is formed by the manifolds of the saddle cycle born from $x^{\mathrm{max}}$,   which become tangent to each other, as described in Section~\ref{sec_homo_cycle}. Nevertheless, the general sequence of bifurcations there is very similar to the one in (\ref{dde_whole}), (\ref{eq_safe}).

\section{Relevance to optimisation}

\label{sec_opt}

\begin{figure*}
\begin{center}
\includegraphics[width=\textwidth]{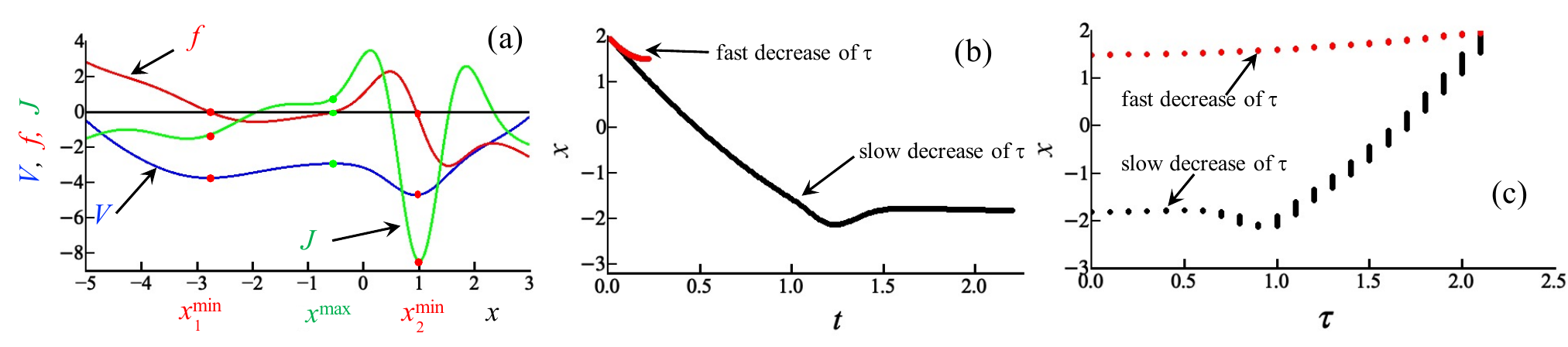}
\caption{Illustration of optimisation by decreasing the delay in system (\ref{dde_whole}), (\ref{eq_safe_low_nf}) where the lowest minimum of $V$ at $x_2^{\textrm{min}}$ is not the flattest or the broadest. Such lowest minimum can be delivered with fast decrease of $\tau$. \\
(a) Functions $V(x)$ (blue line), $f(x)$ (red line) and $J(x)$$=$$f'(x)$ (green line) specified by (\ref{eq_safe_low_nf}). Red/green circles show positions of fixed points at the minima/maximum of $V$. \\
(b)  Solutions $x(t)$ launched from $x(t)$$=$$2$ for $t \in [-2.1,0]$ for $\tau$ decreasing in a step-wise manner  from $2.1$ to zero in steps of $0.1$ of various duration: slower decrease with step $0.1$ (black points), and faster decrease with step $0.01$ (red points).  \\
(c) Points belonging to solutions $x(t)$ shown against the current value of $\tau$ while it decreases as described in (b). Notations are as in (b). }
\label{fig_2well_safe_slow_fast}
\end{center}
\end{figure*}

In the context of optimisation, the delay-induced bifurcations in (\ref{dde_whole})  inform us about the mechanisms behind the removal of the barriers between the local minima of the two-well potential function, which could be potentially applicable to multi-well landscapes as well. Indeed, in the DDE (\ref{dde_whole}) the barriers between the local minima of $V$ are formed by the stable codimension-one manifolds of the saddle fixed point  at  the maximum of $V$, or of the saddle cycle born from this point.  These manifolds are the boundaries of the basins of attractors localised near the minima of $V$. Delay-induced homoclinic bifurcations do not make these fixed points of manifolds disappear, but rearrange manifolds in such a way that they cease to separate different basins and hence to be the barriers. 

The absence of barriers between the local minima and the birth of the large chaotic attractor enables the phase trajectory to visit the vicinities of all minima. Thus, the delay acts by a rough analogy with the random noise in a famous optimisation technique simulated annealing  \cite{Kirkpatrick_simulated_annealing_Sci83}.  

One possible optimisation procedure using the delay $\tau$ as control parameter is illustrated in Fig.~\ref{fig_2well_safe_V_rhs_J_bd}(c). To obtain this figure, Eqs.~(\ref{dde_whole}), (\ref{eq_safe}) were solved 
as  $\tau$ was decreased from some positive value to zero. Namely, at $t$$=$$0$, the system was launched from initial conditions belonging to the large chaotic attractor at $\tau$$=$$3.3$. Then $\tau$ was monotonously decreased in a step-wise manner from $3.3$ to zero with small steps of the size $0.01$. The duration of each step was $50$ time units, during which the phase trajectory approached the attractor closest to the last state from the previous step. The decrease of $\tau$ was almost adiabatic.

With this procedure, at $\tau$$=$$0$ the system ended up at the minimum $x_{1}^{\mathrm{min}}$, at which the modulus of the Jacobian,  $|J|$, is smaller than at $x_{2}^{\mathrm{min}}$ (see Fig.~\ref{fig_2well_safe_V_rhs_J_bd}(a)), and which therefore undergoes AH bifurcation at a larger value of $\tau$ (see Fig.~\ref{fig_2well_safe_V_rhs_J_bd}(b)). This minimum $x_{1}^{\mathrm{min}}$ is not only flatter than $x_{2}^{\mathrm{min}}$, but also broader, due to which the local attractor around $x_{1}^{\mathrm{min}}$ survives for larger values of $\tau$ than the one around $x_{2}^{\mathrm{min}}$. 

Under the adiabatic decrease of $\tau$ within approximately $[1.53,3.3]$, i.e. above the 1st homoclinic, at every instantaneous value of $\tau$ the solution is close to the only attractor corresponding to this value of $\tau$. The latter is either the  one embracing two minima at larger $\tau$, or the one localised near, or at, $x_{1}^{\mathrm{min}}$ at smaller $\tau$. When $\tau$ decreases  below the 1st homoclinic at $\tau$$\approx$$1.53$,  the local attractor around $x_{2}^{\mathrm{min}}$ appears. However, the solution is not affected by this event and remains in the vicinity of $x_{1}^{\mathrm{min}}$ until $\tau$ reaches zero. Since in this example $x_{1}^{\mathrm{min}}$ happens to be the lowest minimum, adiabatic decrease of $\tau$ has demonstrated optimisation as required. 

 In Supplementary Note, Section S--II B, optimisation with adiabatic decrease of $\tau$ is demonstrated for a subtly different example of a system with a two-well potential, in which all  homoclinic bifurcations take a  different form as compared to the ones in (\ref{dde_whole}), (\ref{eq_safe}). To summarise, adiabatic decrease of $\tau$ delivers the flattest and the broadest minimum of the two.

 It is clear that in a general two-well potential such a minimum would not necessarily be the lowest one, and adiabatic decrease of $\tau$ would not result in optimisation. To overcome this issue and to reduce the dependence on the shape of the potential, we can use the fact that larger delay induces chaotic, i.e. random-looking, behaviour forcing the phase trajectory to visit the vicinities of all minima. Then at some sufficiently large value of $\tau$, at some randomly chosen time $t$ the system can be found in one of these vicinities. If at this moment $\tau$ starts to decrease relatively fast, i.e. in a non-adiabatic manner, the phase trajectory would fail to approach any attractor existing at any fixed value of $\tau$. This way, at $\tau$$=$$0$ the system could end up in the same well it was before $\tau$ started to decrease, to whose bottom it would then converge following standard gradient descent (\ref{GDS_nodelay}). Several repetitions of the same experiment with $\tau$  being increased from, and then decreased to, zero would eventually reveal all available minima, whose depths could be compared at the end, and the lowest one would be identified. 

To illustrate this effect, consider (\ref{dde_whole}) with a double-well $V$ and the respective $f$$=$$-V'$ specified as follows
\begin{eqnarray}
\label{eq_safe_low_nf}
V(x)&=&-2 \mathrm{e}^{-2(x-1)^2}- \mathrm{e}^{-0.5(x+3)^2}+0.01(x+1)^4,  \nonumber \\
 f(x)&=&-8 \mathrm{e}^{-2(x-1)^2} (x-1)-  \mathrm{e}^{-0.5(x+3)^2}(x+3)  \nonumber \\
 &-&0.04(x+1)^3.
\end{eqnarray}
The functions $V(x)$, $f(x)$ and $J(x)$$=$$f'(x)$ are shown in Fig.~\ref{fig_2well_safe_slow_fast}(a) by blue, red and green lines, respectively. Red/green circles indicate the positions of the fixed points at the potential minima/maximum. Here,  the flattest and broadest minimum is $x_1^{\textrm{min}}$, but the lowest minimum is $x_2^{\textrm{min}}$. 

First, we launch this system at $t$$=$$0$ from the initial conditions $x(t)$$=$$2$ for $t \in [-2.1,0]$, and $\tau$$=$$2.1$, and start to find the numerical solution $x(t)$ while decreasing $\tau$ at a relatively low rate. Namely, 
after $0.1$ time units, $\tau$ is abruptly decreased to $2.0$ and kept at this value for another $0.1$ time units while the system is being solved. The process continues while $\tau$ is being decreased in a step-wise manner to zero in steps of $0.1$. The resultant solution $x(t)$ is given in Fig.~\ref{fig_2well_safe_slow_fast}(b) by black symbols and demonstrates that with this relatively slow decrease of $\tau$ the system ends up near the flattest and broadest minimum $x_1^{\textrm{min}}$. 

Next, we repeat the process of decreasing $\tau$ step-wise, but make the duration of every step ten times smaller than above, i.e. $0.01$. The resultant solution is shown in  Fig.~\ref{fig_2well_safe_slow_fast}(b) by red symbols, and demonstrates that the system ends up in the well of the lowest minimum $x_2^{\textrm{min}}$. In Fig.~\ref{fig_2well_safe_slow_fast}(c) the same solutions are shown as in (b), but now against the current value of $\tau$ for the convenience of comparison. Thus, the fast decrease of $\tau$ delivers the lowest minimum, which is not the broadest or the flattest.

\section{Discussion and Conclusion}
\label{disc}

The fact of the occurrence of homoclinic bifurcations induced by the increase of delay in (\ref{dde_whole})   with a multi-well potential $V$  could be predicted based on the theorems overviewed in Section~\ref{DDE_simple}. However, the specific  forms of these bifurcations, their dependence on the features of $V$, and their ordering do not  follow from these theorems. 

For a general nonlinear DDE with an arbitrary nonlinearity and an arbitrary dependence on delay, it is usually impossible to predict the delay-induced changes in the behaviour before actually observing this behaviour using numerical analysis. However, we hypothesised 
and verified 
that for a special class of scalar nonlinear models (\ref{dde_whole}), in which the right-hand side 
depends only on the delayed variable and represents the negative of the gradient of a \emph{two-well} potential, it is possible to make some qualitative predictions of the phenomena induced by the increase of the delay.

Specifically, we tested and confirmed our initial hypothesis, that the increase of the delay $\tau$ should lead to a chain of homoclinic bifurcations leading to the disappearance of attractors localised around the minima of $V$, and to the eventual birth of a large attractor embracing  both local minima, which forces the system to visit the vicinity of every minimum, including the global minimum, as time goes by. 

The latter resembles the effect from a random term  within simulated annealing \cite{Kirkpatrick_simulated_annealing_Sci83}, but is achieved in a purely deterministic manner. It also has some similarity with quantum annealing assumed in quantum computers \cite{King_D-wave_techreport_2017}, which is performed thanks to the ability of particles to tunnel the potential barriers between the local minima of energy function. In our case, thanks to the delay, the barrier between the two minima of the cost function disappears, thus enabling the system to freely wander between both minima.  Indeed, the barrier between the minima is ultimately the manifold of the saddle fixed point at the maximum, which separates the basins of attractors localised near the minima. Although after the homoclinic bifurcation neither this manifold, nor the maximum disappear, the manifold stops being the barrier, and hence the barrier ceases to exist. 

 Another important observation is that, for the values of the delay exceeding some threshold, no attractors at finite locations survive in the system, and the phase trajectory tends to infinity from any initial conditions. This effect has been observed numerically for all examples considered, however, it would be useful to verify its validity analytically in the future work. In this context, it would be interesting to understand whether the death of the last attractor at a finite location is linked to reorganisation of manifolds or is caused by other reasons. In the absence of numerical methods for visualisations of the relevant manifolds in DDEs, this matter cannot be resolved by their direct computation at this stage. With this, we are not aware of any relevant theoretical results which could readily suggest a plausible explanation behind this phenomenon. Hopefully, this could be clarified with further development of numerical approaches and/or theory of bifurcations in DDEs. 

Our results show that  the realisation of the particular forms of homoclinic bifurcations depends not only on the parameters of the potential wells and a hump, such as the depth, width and sharpness, but also on the relationships between them. 
Namely, the local attractor around one of the minima may collide either with the maximum itself, or with the saddle cycle born from this maximum. Prior to the disappearance of the local attractor through a homoclinic bifurcation, bistability may either occur, or fail to occur. 

However, the exact nature of homoclinic bifurcations does not seem to change the general sequence of events as $\tau$ grows, which has been confirmed also with a number of additional examples,  including those provided in Supplementary Note.
Thus, we are satisfied with our ability to qualitatively predict the events induced by the increase of time delay in a highly nonlinear system of a special form.

Our results are a pre-requisite to understanding and prediction of the phenomena in general nonlinear delay-differential equations, in which the nonlinear function in the right-hand side depends solely on the delayed variable. 
They will inform building delay systems with prescribed controllable properties in applications.   For example, if one wishes to obtain a system demonstrating a desirable sequence of bifurcations with the increase of delay, one could use the knowledge obtained in order to design the landscape $V$ of (\ref{dde_whole}) with the necessary properties. Namely, one could fine-tune the relative depths, widths and shapes of the individual wells and humps in $V$ in order to ensure a certain order and forms of homoclinic bifurcations. 

A considerable motivation for this study  has been the idea of \cite{Janson_optimization_by_delay_2019} to use delay-induced bifurcations for optimisation. Here we demonstrated how optimisation could be achieved for some  two-well cost functions if one uses the delay as the only control parameter. Further studies will be needed to explore  the possibility to extend this approach to 
cost functions depending on 
many variables.  

~\vspace{-2mm}

\section{Supplementary material}

Supplementary Note contains an overview of the analysis of local dynamics of DDEs around the fixed points (Section S-I), considers an example of a delay system (\ref{dde_whole}) with a double-well potential slightly different from that specified by (\ref{eq_safe}), where a similar, but subtly different, chain of homoclinic bifurcations is observed (Section S-II), and gives three additional examples of double-well potential functions resulting in a similar chain of bifurcations as in examples considered in paper (Section S-III).


\section{Authors' contributions}

NBJ proposed and designed the research, did all numerical simulations described in the paper except for Fig.~\ref{fig_2well_safe_basin}, interpreted the results, and wrote the paper.   CJM performed analytical treatment of a delay differential equation in Section III and in Supplementary Note, did preliminary numerical simulations of a system not illustrated in the paper, which delivered rough initial results, calculated basins of attraction for Fig.~\ref{fig_2well_safe_basin}, and edited the paper. Both authors contributed to the discussions.

\section{Data availability} Data sharing is not applicable to this article as no new data were created or analysed in this study.

\section{Acknowledgements}

The authors are grateful to Dmitri Tseluiko for helpful discussions about Shilnikov's theorem, and for help in preparing Figures  \ref{fig_homocl_2d-3d_2well_right_cycle}(k)-(o); and to Alexander Balanov for helpful discussions and feedback on the draft of this manuscript.  C.J.M. was supported by EPSRC (UK) during his PhD studies in Loughborough University,  grant EP/P504236/1.

\end{document}


\preprint{AIP/123-QED}

\title{Delay-induced homoclinic bifurcations in modified gradient bistable systems and their relevance to optimisation\\ \vspace{3mm}
SUPPLEMENTARY NOTE}


\author{Natalia~B.~Janson}
\email[E-mail: ]{N.B.Janson@lboro.ac.uk}
\author{Christopher~J.~Marsden}
\affiliation{Department of Mathematics, Loughborough University, Loughborough LE11 3TU, UK}

\maketitle

\setcounter{equation}{0}
\setcounter{figure}{0}
\setcounter{table}{0}
\setcounter{page}{1}
\makeatletter
\renewcommand{\theequation}{S\arabic{equation}}
\renewcommand{\thefigure}{S\arabic{figure}}
\renewcommand{\bibnumfmt}[1]{[S#1] }
\renewcommand{\citenumfont}[1]{S#1}
\renewcommand\thesection{S--\Roman{section}}

\section{Analysing dynamics near fixed points in a delay differential equation}

Here we show how the standard analytical tools can be applied to predict 
the local behaviour of the delay-differential equation (DDE) of the following form 
\begin{equation}
\label{dde_whole}
\dot{x}=f(x_{\tau}), \quad f(z)=-\frac{\mathrm{d} V(z)}{\mathrm{d}  z},
\end{equation}
where $x, f, V, z$ $\in$ $\mathbb{R}$, $x_{\tau}$$=$$x(t-\tau)$, $V(z)$ is some landscape function, and $f(z)$ is a twice differentiable function. Specifically, by performing linear stability analysis, we will reveal at what values of time delay $\tau$ the fixed points at the minima of $V(z)$ change from stable nodes to stable focuses and then undergo Andronov-Hopf bifurcation. 
Doing so will allow us to predict at what $\tau$ around the given fixed point a stable periodic oscillation could be observed. 

We also analyse the stability of the fixed points located at the maxima of $V(z)$. Although these points remain unstable for any positive $\tau$ and thus unobservable in an experiment, perhaps counterintuitively, the parameters characterising their instability affect the observable behaviour of the system. For example, it is possible for the manifolds of an unstable fixed point of a saddle-focus type to close and to form a homoclinic loop.  As as the parameter of the system changes, this loop breaks down and can give birth to a new attractor, which will extend well beyond the close vicinity of the relevant point. According to Shilnikov's theorem,  this new attractor can be either periodic or chaotic, and its nature is determined by the local properties of the respective unstable point$^{55,56,}$\cite{Shilnikov_on_Shilnikov_theorem_url07,Afraimovich_heritage_of_Shilnikov_RCD14}.

We start from performing linear stability analysis of the fixed points ${x}^{*}$ of (\ref{dde_whole}), which satisfy
\begin{equation}
{f}({x}^{*}) = {0}.
\label{eqnequil}
\end{equation}
Unlike a system of ordinary differential equations (ODEs), a DDE is an infinite-dimensional dynamical system because in order to solve such an equation, as an initial condition one needs to specify a function on an interval, namely, $x(t)$$=$$\varphi(t)$ on $t \in [-\tau,0]$, which contains infinitely many points. With this, if we wish to represent the dynamics of a DDE in the phase space by analogy with what is done for ODEs, we need to appreciate that the full state of a DDE at time $t$ would a function on an interval  $[t-\tau, t]$, and therefore the dimension of the respective phase space is infinity \cite{Otto_DDE_PTRS19}. 
As a single state of the DDE, the fixed point can be viewed as a constant function ${x}(t)$$=$${x}^{*}$ on an interval $[t-\tau, t]$, which stays the same at all times $t$. By analogy with the linear stability analysis of ODEs, we need to slightly perturb the system away from the fixed point. So we introduce 
\begin{equation*}
{x} = {x}^{*} + {X}, \quad \big|{X}\big| \ll 1,
\end{equation*}
where  ${X}(t)$ is the perturbation of the fixed point ${x}^{*}$. 
Given that $\dot{{x}} = \dot{{X}}$, substitute the latter into (\ref{dde_whole}) to obtain 
\begin{equation*}
\dot{{X}} = {f}({x}^{*}+{X}_{\tau}),
\end{equation*}
where ${X}_{\tau}(t)$$=$${X}(t-\tau)$. Because ${X}$ is small, $\big|{X}_{\tau}\big|$$\ll$$1$,  and we can approximate the above equation by expanding ${f}$ in Taylor series and keeping the linear terms only
\begin{equation*}
\dot{{X}} \approx {f}({x}^{*}) + {J}{X}_{\tau},
\end{equation*}
where ${J}$$=$$\frac{\mathrm{d} f}{\mathrm{d} z}$ at $z$$=$$x^*$. 
With account of (\ref{eqnequil}), we obtain the so-called {\it linearized equation}, which holds as long as ${X}(t)$ stays small
\begin{equation}
\dot{{X}} = {J} {X}_{\tau}.
\label{eqnlin}
\end{equation}
We assume that the linear DDE (\ref{eqnlin}) has exponential solutions, i.e. 
\begin{equation*}
X(t) = Ae^{\lambda t}, \quad A=\mathrm{const}, 
\end{equation*}
which on substitution back into (\ref{eqnlin}) gives
\begin{equation}
\label{eqnmddet}
\lambda= J e^{-\lambda\tau}.
\end{equation}
Equation (\ref{eqnmddet}) is called the characteristic equation, and its roots $\lambda$ are called the eigenvalues of the fixed point. Eq.~(\ref{eqnmddet})  has a countably infinite number of generally complex roots.  The eigenvalue with the largest real part is called the leading eigenvalue. If the solution of the characteristic equation has only negative real parts, then the respective fixed point is stable. If there are eigenvalues with positive real parts, then the fixed point is unstable.  If the real part of the leading eigenvalue is zero, the stability cannot be deduced to linear order. 
Although an infinite number of eigenvalues could potentially present a challenge in the stability analysis, it has been proved that there are only a finite number of eigenvalues to the right of any vertical line in the complex plane \cite{Hale_DDE_intro_book93}. This means, that the number of eigenvalues with a positive real part can only be finite, so to determine the stability of a fixed point, we need to consider only a finite number of values. 

\begin{figure}
\begin{center}
\includegraphics[width=0.3\textwidth]{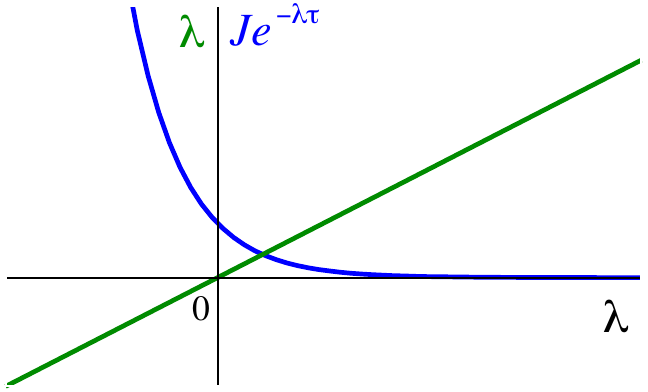}
\caption{Illustration of solution to Eq.~(\ref{eqnmddet}). Figure shows the graphs of $\lambda$ (green line) and of $J e^{-\lambda\tau}$ for positive $\tau$ and $J$ (blue line).  These two curves are guaranteed to intersect at some positive and real value of $\lambda$, which means that the local maxima of the landscape $V(z)$ in (\ref{dde_whole}) are always unstable. }
\label{eig_sketch_saddle}
\end{center}
\end{figure}

Note, that the fixed points corresponding to the maxima of $V(z)$ are unstable at any $\tau$$ \ge$$0$. Indeed,  at a maximum of $V(z)$, ${J}$$=$$\frac{\mathrm{d} f}{\mathrm{d} z}$$=$$-\frac{\mathrm{d}^2 V}{\mathrm{d} z^2}$$>$$0$.   With a positive $J$,  Eq.~(\ref{eqnmddet}) always has a real positive root $\lambda$ (see Fig.~\ref{eig_sketch_saddle} 
illustrating the solution). 

\begin{figure}
\begin{center}
\includegraphics[width=0.45\textwidth]{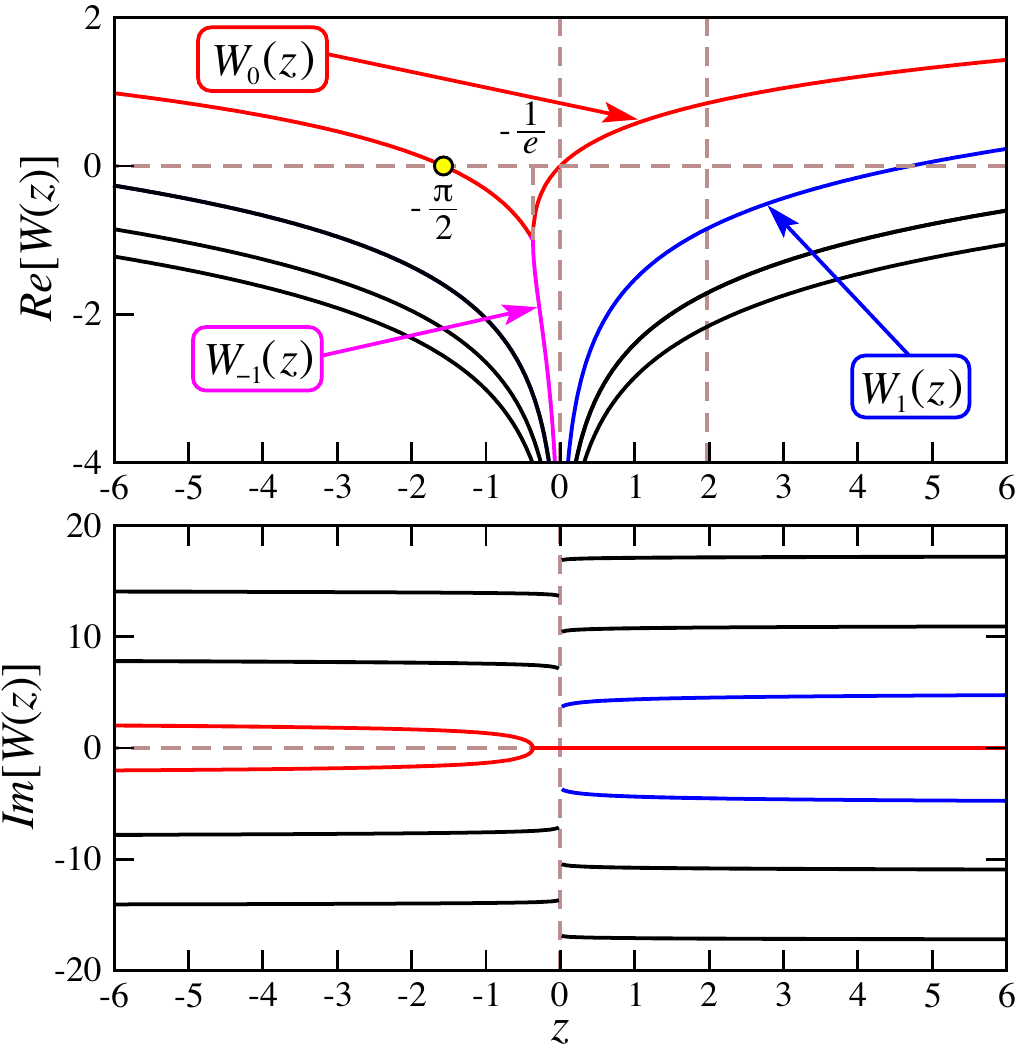}
\caption{Real (upper panel) and imaginary (lower panel) parts of Lambert Function $W(z)$, assuming that $z$ takes only real values. Red line shows the principal branch $W_0(z)$, magenta line shows $W_{-1}(z)$, and blue line shows $W_{1}(z)$.}
\label{fig_lambert}
\end{center}
\end{figure}

The solutions $\lambda$ of (\ref{eqnmddet}) can be expressed in terms of the Lambert function $W(z)$, also called the product log function. Lambert function is a complex-valued function of a complex argument defined as an inverse of the function $h(z)$$=$$ze^z$, where $z, h$ $\in$ $\mathbb{C}$ \cite{Corless_on_Lambert_function_ACM96,Dubinov_Lambert_function_JPP05}. The function $W(z)$ is formed of solutions $w$ of the equation 
\begin{equation}
we^{w} = z, 
\label{eqnlam}
\end{equation}
so that for every value of $z$ there is a countable infinity of the values of $w$. In applications to the stability of DDEs, 
only real arguments $z$ of $W(z)$ are used, which means that only the values of  $W(z)$ corresponding to the cross-section along $Im[z]$$=$$0$ are relevant to the problem. From now on, we will assume that $z$ are real. The function $W(z)$ with 
$z$$\in$$\mathbb{R}$ has countably many branches $W_{k}$, $k = 0, \pm 1, \pm 2, \ldots$, and $W_{0}$ is called the principal branch. In Fig.~\ref{fig_lambert}, 
real (upper panel) and imaginary (lower panel) parts of $W(z)$ are shown as a function of $z$.

The solutions $\lambda$ of (\ref{eqnmddet}) are equal to 
\begin{equation}
\label{eig_lambert}
\lambda_k = \frac{W_{k}(J \tau)}{\tau}=\frac{JW_{k}(z)}{z}, 
\end{equation}
where $W_{k}$ is the $k$-th branch of $W(z)$. Note, that in (\ref{eig_lambert}), 
for $z>-\frac{1}{e}$ the principal branch $W_0(z)$ takes real values only. In Fig.~7 eigenvalues $\lambda_k$ are illustrated as functions of $(J\tau)$ for the fixed points at local maxima of the landscape $V$ of (\ref{dde_whole}) with $J$$>$$0$. As confirmed by Fig.~7, such fixed points are always unstable because one of their eigenvalues is real and positive at all positive values of $\tau$. Note, that at $\tau$$=$$0$ these fixed points have only one positive real eigenvalue equal to $J$. Indeed, at $\tau$$=$$0$ (3) becomes a first order ODE, in which fixed points have only one real eigenvalue equal to the derivative $J$ of the right-hand side. Fig.~6 shows eigenvalues as functions of $(|J| \tau)$ for the fixed points at local minima of  $V$ with $J$$<$$0$. One can see that for small $\tau$ these fixed points are stable, but as $\tau$ increases, they become unstable. 

Next, we calculate the values of $\tau$, at which Andronov-Hopf (AH) bifurcation occurs for the given fixed point. If a stable fixed point undergoes   AH bifurcation, the point becomes unstable and possibly a stable periodic solution is born from it, thus giving rise to oscillatory behaviour of the system. The first condition of AH bifurcation is that the real part of $W_{0}(J\tau)$ must be equal to zero. 
Given $J$, we need to determine the respective value of $\tau$ 
at which (\ref{eqnmddet}) has purely imaginary roots, i.e. $\lambda$$=$$\pm \beta i$ with $\beta$$\in$$\mathbb{R}$. Substituting this into (\ref{eqnmddet}) gives
\begin{equation*}
\beta i - J e^{-\beta i \tau} = 0.
\end{equation*}
This can be written as
\begin{equation*}
\beta i - J[(\cos(\beta \tau) - i\sin(\beta \tau)] = 0.
\end{equation*}
Separating real and imaginary parts gives
\begin{eqnarray}
\cos(\beta \tau) &=& 0, \label{eqnreal}\\
\beta + J \sin(\beta \tau) &=& 0 \label{eqnimag}. 
\end{eqnarray}
Equation (\ref{eqnreal}) has infinitely many solutions in the form of conjugate pairs,
\begin{equation}
\label{AH_im}
\beta = \frac{\left(\frac{\pi}{2} \pm n\pi\right)}{\tau}, \hspace{1cm} n=0,1,2,\ldots ,
\end{equation}
which we then substitute into (\ref{eqnimag}) to obtain
\begin{equation}
\tau = -\frac{\left(\frac{\pi}{2}+n\pi\right)}{J (-1)^{n}}, \hspace{1cm} n=0,1,2,\ldots ,
\label{eqnhb}
\end{equation}
which will have the same $\tau$ for both components of the conjugate pair.  Assume that we consider the local minima of the landscape $V(z)$, at which $J$$<$$0$.  Also, in this work we assume that $\tau$ can only be positive, and to ensure that this is the case in (\ref{eqnhb})  $n$ must be \emph{even}. The first Andronov-Hopf bifurcation occurs for the smallest even value of $n$, therefore we take $n$$=$$0$ to obtain 
\begin{equation}
\tau_{AH} = -\frac{\pi}{2J}=\frac{\pi}{2|J|}.
\label{eqnhbcrit}
\end{equation}
This result is consistent with the earlier predictions, obtained with a very different approach, of the existence of periodic solutions in DDEs (4) for a special form of $g(z)$  as discussed in Sec.~IIA. 
At larger even values of $n$ in (\ref{eqnhb}),  more complex conjugate pairs of eigenvalues cross the imaginary axis from below. 

The second criterion to be satisfied for an Andronov-Hopf bifurcation to occur is \cite{Hale_FDE_book71,Hale_DDE_intro_book93}
\begin{equation}
\label{AH2}
\frac{d Re(\lambda)}{d\tau}\Big|_{\tau = \tau_{AH}} > 0.
\end{equation}
To verify this, differentiate both parts of (\ref{eqnmddet}) with respect to $\tau$ to obtain
\begin{equation}
\label{la1}
\frac{d \lambda}{d \tau} = -Je^{-\lambda \tau} \left( \lambda + \tau \frac{d \lambda}{d \tau} \right) .
\end{equation}
Assume that $\lambda$$=$$\alpha$$+$$i \omega$ with $\alpha$, $\omega$ $\in$ $\mathbb{R}$, and substitute into (\ref{la1}). Separating real and imaginary parts gives
\begin{eqnarray}
\frac{d \alpha}{d \tau} =& - & Je^{-\alpha \tau} \cos(\omega\tau)\left(\alpha + \tau \frac{d \alpha}{d \tau} \right) \nonumber\\
&-& Je^{-\alpha \tau}\sin(\omega\tau)\left( \omega + \tau \frac{d \omega}{d \tau}\right),\nonumber\\
\frac{d \omega}{d \tau} =& - &Je^{-\alpha \tau} \cos(\omega\tau)\left(\omega + \tau \frac{d \omega}{d \tau} \right) \nonumber\\ 
&+& Je^{-\alpha \tau}\sin(\omega\tau)\left( \alpha + \tau \frac{d \alpha}{d \tau}\right).
\label{eqnadash}
\end{eqnarray}
Eq.~(\ref{AH2}) implies 
\begin{equation}
\frac{d\alpha}{d\tau}\Big|_{\tau=\tau_{AH}} > 0.
\end{equation}
At AH bifurcation $\alpha$$=$$0$ and $\omega$$=$$\beta$. By setting in (\ref{AH_im})  $n$$=$$0$ and substituting $\tau$ from  (\ref{eqnhbcrit}) we obtain 
\begin{eqnarray}
\alpha &=& 0,\nonumber\\
\omega &=& \frac{\pi}{2\tau_{AH}} = -J.
\end{eqnarray}
Substituting these values into (\ref{eqnadash}) gives
\begin{eqnarray}
\frac{d \alpha}{d \tau} &=& J^{2} + \frac{\pi}{2}\frac{d\omega}{d\tau},\nonumber\\\nonumber\\
\frac{d \omega}{d \tau} &=& -\frac{\pi}{2}\frac{d\alpha}{d\tau}.\nonumber\\
\end{eqnarray}
Combining these equations to exclude $\frac{d\omega}{d\tau}$ gives
\begin{equation}
\frac{d \alpha}{d \tau} = \frac{4J^{2}}{(4+\pi^{2})} > 0.
\end{equation}
So we have verified both criteria for the occurrence of AH bifurcation, in which the fixed point loses stability. In the numerical simulations performed here, the bifurcation is always supercritical and leads to the birth of a stable limit cycle.

Equation (\ref{eqnhb}) can also be used to find the values of $\tau$ at which the fixed points at the maxima of the landscape $V(z)$ undergo AH bifurcation.  In this case $J$$\geq$$0$, therefore, to ensure the positivity of $\tau$, $n$ must be \emph{odd}. Again choosing the smallest value of $n$, $n$$=$$1$, gives
\begin{equation}
\tau_{AH}^s = \frac{3\pi}{2J}.
\end{equation}
Such fixed points would remain unstable for any $\tau$. However, as a result of this bifurcation, a saddle periodic orbit could be born from the fixed point, together with its stable and unstable manifolds, which will affect the dynamics of the system. Further complex conjugate pairs will cross the imaginary axis for odd values of $n$ in (\ref{eqnhb}).

This theory can be extended to equations of the form (2) (or (\ref{dde_whole})) with an $n$-dimensional vector $x$. The approach will be similar, however, an $n$$ \times $$n$ Jacobian matrix will be involved, resulting in an order-$n$ quasi-polynomial for the characteristic equation, which can be analysed to deduce the relevant bifurcations.

\begin{figure}
\begin{center}
\includegraphics[width=0.4\textwidth]{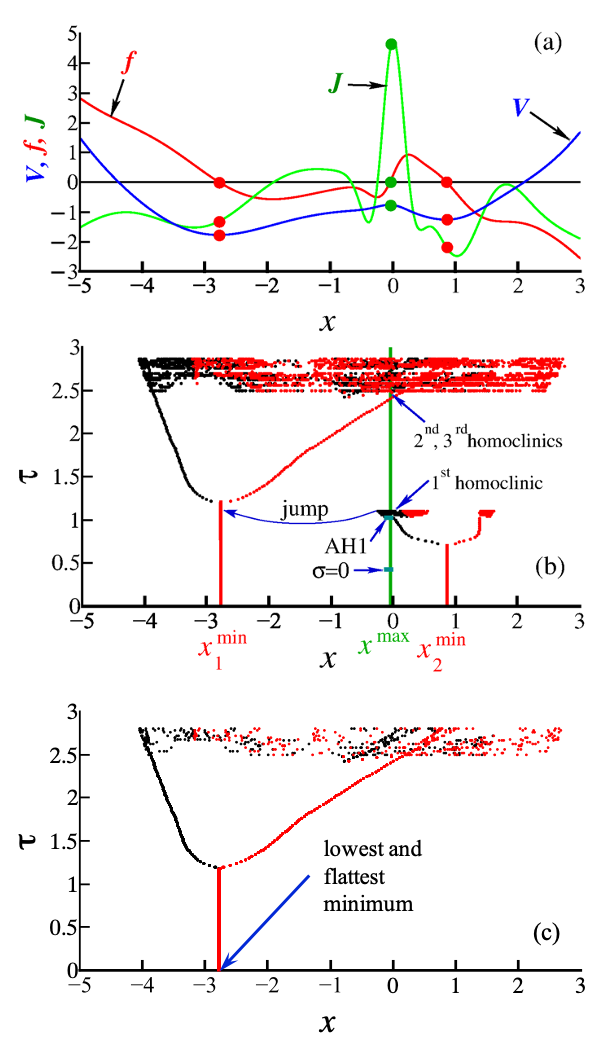}
\caption{ (a) Functions $V(x)$ (blue line), $f(x)$ (red line) and $J(x)$$=$$f'(x)$ (green line) specified by (\ref{eq_after_AH}). Red/green circles show positions of fixed points at the minima/maximum of $V$. \\
(b) Bifurcation diagram of (2) (or (\ref{dde_whole})),  (\ref{eq_after_AH}). Local minima/maxima of attractors are shown by black/red dots. Fixed points at the minima of $V$ are shown by red vertical lines for $\tau$ at which they are stable, and the saddle-focus at the maximum of $V$ is shown by green vertical line. 
As $\tau$ varies,  homoclinic  orbits are formed by the manifolds of the saddle cycle around $x^{\mathrm{max}}$ born at the point AH1. \\
(c) Demonstration of optimisation. Local maxima (red dots) and minima (black dots) of solutions to (\ref{dde_whole}), (\ref{eq_after_AH}), which are obtained as $\tau$ \emph{slowly} decreases from $2.8$ to zero.  The solution spontaneously settles down at the lowest,  flattest and broadest minimum at $x_1^{\mathrm{min}}$. Compare with (a) and (b).   }
\label{fig_2well_after_AH_V_rhs_J_bd}
\end{center}
\end{figure}

\section{Homoclinic orbit to a saddle cycle in DDE (\ref{dde_whole})  with two-well potential}

\label{sec_hom_cycle}

In an example below, we illustrate a chain of global homoclinic bifurcations in a DDE of the form Eq.~(\ref{dde_whole})  (same as Eq.~(2) in main paper) with a two-well potential $V$, in which all homoclinic bifurcations come in a different form as compared to the ones in  Eqs.~(2), (9). 
Namely, in (2), (9) all three global bifurcations occurring as the delay $\tau$ grows    consist in the formation of homoclinic loops of the saddle-focus fixed point at the maximum of $V$. However, here some of these  manifest themselves in  the formation of homoclinic orbits to the saddle \emph{ cycle}. These homoclinic orbits  arise as the manifolds 
of the saddle cycle, born from the saddle fixed point at the maximum,  become tangent to each other. Although the latter bifurcations are somewhat more complex than the ones involving the fixed point, they can nevertheless be routinely expected to occur in the systems of the form (2).

To illustrate the homoclinic bifurcations involving the manifolds of the saddle cycle, we subtly modify the double-well landscape $V$  of (9) to make the maximum of $V$ considerably sharper, by adding an extra term to $V$ as specified below 
\begin{eqnarray}
\label{eq_after_AH}
V(x)=&- &\frac{1}{2} \mathrm{e}^{-2(x-1)^2}- \mathrm{e}^{-0.5(x+3)^2} \nonumber\\ 
& + & 0.01(x+1)^4 
+0.2\mathrm{e}^{-10x^2},  \nonumber \\
 f(x)=&- & 2 \mathrm{e}^{-2(x-1)^2} (x-1)-  \mathrm{e}^{-0.5(x+3)^2}(x+3)  \nonumber\\
 &-&  0.04(x+1)^3+4 \mathrm{e}^{-10x^2}x. 
\end{eqnarray}
With this, the locations, depths and shapes of both minima stay roughly the same as in (9) to make the results comparable. 
The functions $V(x)$, $f(x)$ and $J(x)$$=$$f'(x)$ are shown in Fig.~\ref{fig_2well_after_AH_V_rhs_J_bd}(a) by blue, red and green lines, respectively.

This way, in  (\ref{eq_after_AH}) the Jacobian  of the saddle-focus fixed point at the maximum 
$x^{\mathrm{max}}$$\approx$$-0.0420246$ is $J(x^{\mathrm{max}})$$\approx$$4.54$ and is  much larger than $0.49$ in (9). As a result, its AH bifurcation occurs at a much smaller value of  $\tau$$\approx$$1.037783$ as compared to $9.61$ in (2), (9). 

\subsection{Bifurcations}

Similarly to 
(2), (9), 
in Eqs.~(2), 
(\ref{eq_after_AH}) the increase of $\tau$ leads to a sequence of three homoclinic bifurcations illustrated by the bifurcation diagram in Fig.~\ref{fig_2well_after_AH_V_rhs_J_bd}(b)  (compare with Fig.~8(b)).

\begin{figure}
\begin{center}
\includegraphics[width=0.4\textwidth]{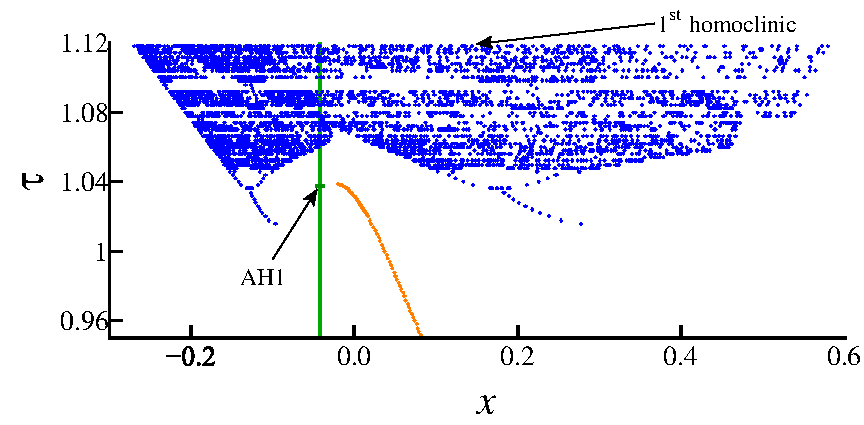}
\caption{ A segment of the bifurcation diagram in Fig.~\ref{fig_2well_after_AH_V_rhs_J_bd}(b)  illustrating bistability, hysteresis, and the 1$^{\textrm{st}}$ homoclinic bifurcation in (2) (or (\ref{dde_whole})),  (\ref{eq_after_AH}) eliminating the localised attractor 
inside the right well of $V$  
(compare with Fig.~5). The version of the homoclinic bifurcation here  manifests itself in the  homoclinic orbit formed by the  tangency of the manifolds of the saddle cycle around the potential maximum $x^{\mathrm{max}}$. \\
Green line shows $x^{\mathrm{max}}$. Dots show local minima of attractors: orange dots show   limit cycle  born via the first AH bifurcation  from $x_{2}^{\mathrm{min}}$ (outside the range of this figure), and blue dots show attractor developed from the homoclinic  orbit of the saddle cycle via 1$^{\textrm{st}}$ homoclinic bifurcation as  $\tau$ {\it decreases} from $1.118$.  AH1 is the first AH bifurcation of  $x^{\mathrm{max}}$, which gives rise to the saddle cycle  together with its manifolds at $\tau$$\approx$$1.118$. }
\label{fig_2well_after_AH_hom_right_bd}
\end{center}
\end{figure}

\begin{figure}
\begin{center}
\includegraphics[width=0.5\textwidth]{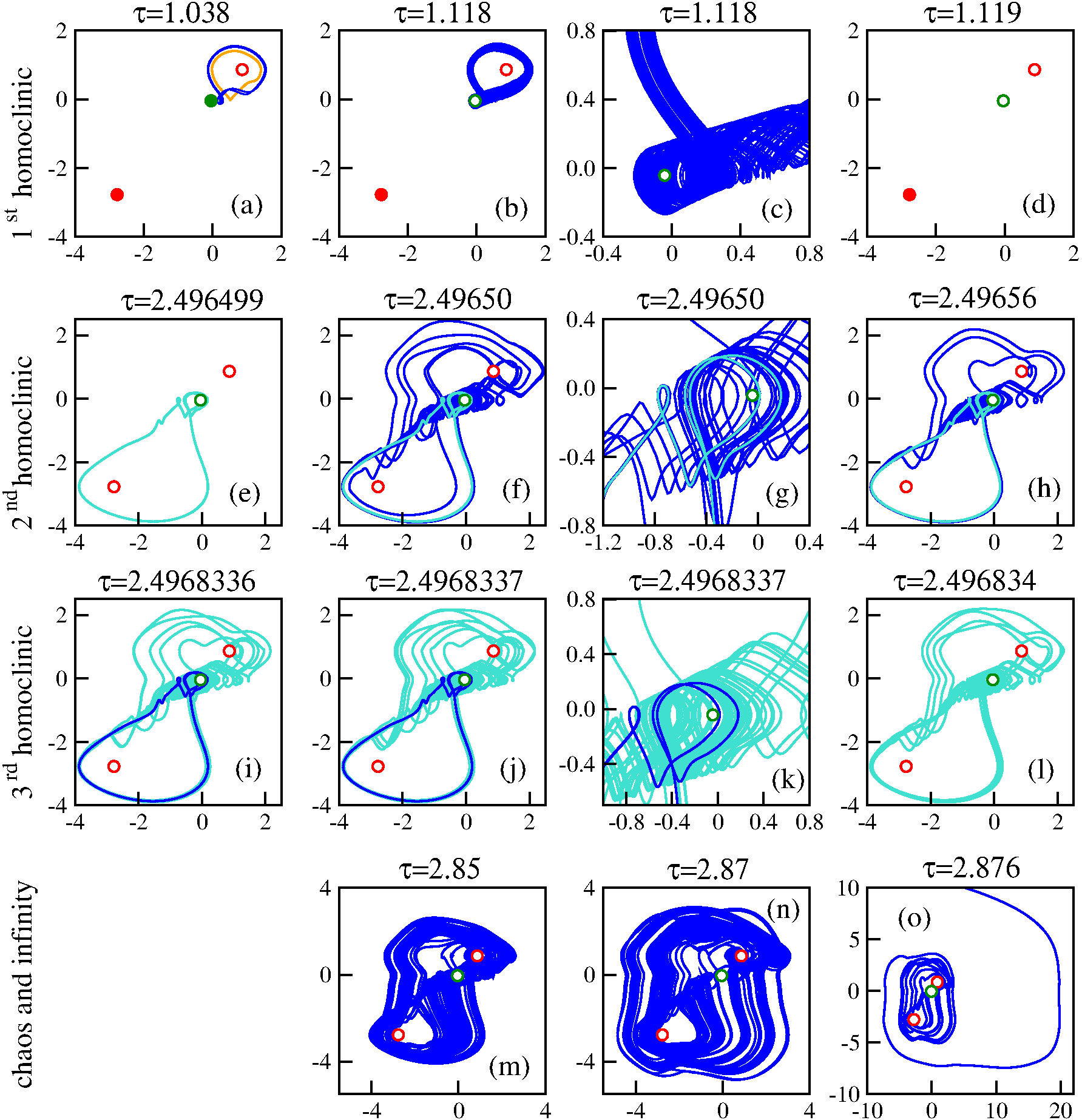}
\caption{Phase portraits of  (2) (or (\ref{dde_whole})), (\ref{eq_after_AH}) at various $\tau$ illustrating bifurcation diagrams in Figs.~\ref{fig_2well_after_AH_V_rhs_J_bd} and \ref{fig_2well_after_AH_hom_right_bd} with $\tau$ given in panels. Notations are: fixed points $x_{1,2}^{\mathrm{min}}$ (red circles) and $x^{\mathrm{max}}$ (green circle); filled/empty circles indicate points below/above AH bifurcations;   attractors (blue, turquoise and orange lines). 
{\bf 1$^\textrm{st}$ row:} (a) Before, (b)--(c) at, and (d) after the 1st homoclinic bifurcation. 
(a) Three coexisting attractors: stable fixed point $x_{1}^{\mathrm{min}}$ (red filled circle) and two limit cycles around $x_{2}^{\mathrm{min}}$ (orange and blue lines). (b)--(c) Small chaos born from the homoclinic orbit to  the saddle cycle around $x^{\mathrm{max}}$ (blue line) (compare with Fig.~4(c)) 
coexists with stable fixed point $x_{1}^{\mathrm{min}}$. (d) Fixed point $x_{1}^{\mathrm{min}}$ is the only attractor. 
{\bf 2$^\textrm{nd}$ row:}
 (e) Before, (f)--(g) at, and (h) after   the 2nd homoclinic bifurcation. 
(e) Small cycle around $x_{1}^{\mathrm{min}}$ is the only attractor (turquoise line). (f)--(g) Birth of large chaos (blue line) from the homoclinic orbit to the saddle cycle around $x^{\mathrm{max}}$ . (h) Large chaos (blue line) coexists with the small limit cycle  around $x_{1}^{\mathrm{min}}$ (turquoise line).
 {\bf 3$^\textrm{rd}$ row:}
 (i) Before, (j)--(k) at, and (l) after the 3rd homoclinic bifurcation. (i)--(k) Small cycle (blue line) coexists with large chaos (turquoise line). (j)--(k) Small attractor (blue line) almost at (just before) the 3rd homoclinic bifurcation. (l) Large chaos is the only attractor. 
 {\bf 4$^\textrm{th}$ row:}
 (m) Chaos embracing all fixed points at large $\tau$, (n) chaotic attractor grows in size, and (o) trajectory goes to infinity at $\tau$ above some critical value.  }
\label{fig_2well_after_AH_pp}
\end{center}
\end{figure}

Specifically, as $\tau$ grows from zero, the sequence of events is initially the same as in (2), (9). 
 At $\tau$$\approx$$0.705$ a stable cycle is born through AH bifurcation from $x_2^{\mathrm{min}}$, which grows in size with $\tau$ (orange dots in Fig.~\ref{fig_2well_after_AH_hom_right_bd} and orange line in Fig.~\ref{fig_2well_after_AH_pp}(a)). However, at $\tau$$\approx$$1.038$ this cycle disappears via a mechanism not yet clear. 
In a narrow range $\tau$$\in$$[1.02,1.038]$ two attractors coexist around $x_2^{\mathrm{min}}$ (compare Figs.~\ref{fig_2well_after_AH_pp}(a) and 4(f)),  but this multistability does not affect considerably the general picture of bifurcations predicted for a system with a double-well landscape $V$.

The 1st homoclinic bifurcation, which comes in a different form as compared to (2), (9), is illustrated in more detail by a segment of the bifurcation diagram in Fig.~\ref{fig_2well_after_AH_hom_right_bd} (compare with 
Fig.~5) and by the phase portraits in  Fig.~\ref{fig_2well_after_AH_pp}(a)--(d).  To explain this bifurcation, it is convenient to consider events as $\tau$ \emph{decreases}. 
The first small homoclinic  orbit occurs 
at $\tau$$\approx$$1.118$  when the manifolds of the saddle cycle become tangent to each other. As $\tau$ is decreased, 
 these manifolds are no longer tangent, but instead intersect each other transversally, and a chaotic attractor exists in the region of this intersection
(compare   Fig.~\ref{fig_2well_after_AH_pp}(b)--(c) with Fig.~4(c)). The structure of manifolds in this situation is quite complicated and difficult to visualise even schematically, and to get a better understanding of it the reader might benefit from a more detailed description in$^{20}$. With the further reduction of $\tau$, chaos undergoes an inverse sequence of period-doubling bifurcations transforming it into a limit cycle (compare blue lines in Fig.~\ref{fig_2well_after_AH_pp}(a) and Fig.~4(f)). The latter cycle disappears at $\tau$$\approx$$1.02$.

Now, if one increases $\tau$  beyond the value of the 1st homoclinic $1.118$, all of the local attractors to the  right  of  $x^{\mathrm{max}}$, i.e. near $x_2^{\mathrm{min}}$,  cease to exist, and the system has a single local attractor around $x_1^{\mathrm{min}}$ (red filled circle in Fig.~\ref{fig_2well_after_AH_pp}(d) and cyan line in Fig.~\ref{fig_2well_after_AH_pp}(e)). Specifically, if at $\tau$ just above $1.118$ the initial conditions are set on or near the just disappeared local attractor around $x_2^{\mathrm{min}}$, the phase trajectory swiftly leaves this region and converges to the stable fixed point $x_1^{\mathrm{min}}$ in what might feel like a ``jump" as indicated in Fig.~\ref{fig_2well_after_AH_V_rhs_J_bd}(b).

At some higher value of $\tau$$\approx$$2.49650$, the 2nd (large) homoclinic orbit is formed by  the tangency of  another pair of manifolds of the saddle cycle around  $x^{\mathrm{max}}$, which gives birth to a large chaotic attractor embracing all three fixed points (blue line  in Figs.~\ref{fig_2well_after_AH_pp}(f), (h)).  This can be interpreted as follows: at $\tau$ slightly above the bifurcation value, these manifolds intersect transversally and ensure the existence of a large chaotic attractor in the same region of the phase space. As $\tau$ decreases to approach the bifurcation value, the manifolds cease to intersect transversally and instead only touch each other, thus forming a non-robust homoclinic orbit to the saddle cycle. At $\tau$  below the bifurcation value, the manifolds no longer intersect, there is no large homoclinic orbit and no large chaotic attractor. 

In a narrow range $\tau$$\in$$[2.49650,2.4968337)$ two attractors coexist: the only remaining local attractor around $x_1^{\mathrm{min}}$ (blue line  in Fig.~\ref{fig_2well_after_AH_pp}(i)) and the large attractor (cyan  line  in Fig.~\ref{fig_2well_after_AH_pp}(i)). 

At  $\tau$$\approx$$2.4968337$ the 3rd homoclinic  bifurcation occurs leading to the disappearance of  
the last of the local attractors.  At higher $\tau$, the only remaining attractor is the large chaos (cyan line in Fig.~\ref{fig_2well_after_AH_pp}(l)), which survives until $\tau$$\approx$$2.87$
(blue line in Figs.~\ref{fig_2well_after_AH_pp}(m)--(n)). Like in (2), (9), 
as $\tau$ reaches a certain threshold (here $2.876$), even this large attractor vanishes and the phase trajectory goes to infinity (blue line in Fig.~\ref{fig_2well_after_AH_pp}(o)).

\subsection{Optimisation}

One possible optimisation procedure is illustrated in Fig.~\ref{fig_2well_after_AH_V_rhs_J_bd}(c). To obtain this figure, Eqs.~(2) (or (\ref{dde_whole})), (\ref{eq_after_AH}) were solved with the parameter $\tau$ being \emph{slowly} decreased in a step-wise manner from the value $2.8$  to zero, after the system was launched from initial conditions belonging to the chaotic attractor at $\tau$$=$$2.8$. 
The procedure was the same as the one described in Section~V of the main paper. 

The system ended up at the minimum $x_{1}^{\mathrm{min}}$, at which the modulus of the Jacobian,  $|J|$, is smaller than at $x_{2}^{\mathrm{min}}$ (see Fig.~\ref{fig_2well_after_AH_V_rhs_J_bd}(a)), and which therefore undergoes AH bifurcation at a larger value of $\tau$ (see Fig.~\ref{fig_2well_after_AH_V_rhs_J_bd}(b)). This minimum $x_{1}^{\mathrm{min}}$ is not only flatter than $x_{2}^{\mathrm{min}}$, but also broader, due to which the local attractor around $x_{1}^{\mathrm{min}}$ survives for larger values of $\tau$ than the one around $x_{2}^{\mathrm{min}}$. For this reason, for the current shape of $V$ adiabatic decrease of $\tau$ delivered the global minimum. 

An example with the lowest minimum not being the flattest or belonging to the broadest well is considered in the main paper, Section~V, where it is shown how such a minimum could be found with the fast decrease of $\tau$.

\section{Additional examples with similar behaviour}

In the main paper we provided a detailed illustration of a chain of homoclinic bifurcations occurring in a system of the form (2) (or (\ref{dde_whole})) with a double-well potential described by (9). In Section~\ref{sec_hom_cycle}  we illustrated a similar chain  with a slightly different form of  homoclinic bifurcations in a subtly different system whose potential function is given by (\ref{eq_after_AH}). In addition, we observed qualitatively the same bifurcations in  (2) with slightly different versions of double-well potentials $V$,  which we do not illustrate here. Three  of these examples are given below. 

\begin{eqnarray}
\label{eq_extra}
V(x)=&- & \frac{1}{2} \mathrm{e}^{-2(x-1)^2}- \mathrm{e}^{-0.5(x+3)^2} \nonumber\\
&+& 0.01(x+1)^4 +0.07\mathrm{e}^{-10x^2},   \\ 
V(x)=&-&\frac{1}{2} \mathrm{e}^{-2(x-1)^2}- \mathrm{e}^{-0.5(x+3)^2}\nonumber\\
& + &0.1(x+1)^2 
 \\
 V(x)=&-& \frac{1}{40} x^4 -\frac{0.05}{3} x^3 + \frac{1}{4} x^2  
 \end{eqnarray}